\documentclass[12pt]{article}

\usepackage{amssymb}
\usepackage[dvips]{graphicx}

\newcommand {\beq}{\begin{equation}}
\newcommand {\eeq}{\end{equation}}
\newcommand {\beqa}{\begin{eqnarray}}
\newcommand {\eeqa}{\end{eqnarray}}
\newcommand {\n}{\nonumber \\}
\newcommand {\tr}{\mbox{tr}}
\newcommand {\Tr}{\mbox{Tr}}
\newcommand {\trs}{\mbox{\scriptsize tr}}

\newcommand {\ee}{\mbox{e}}
\newcommand {\ees}{\mbox{\scriptsize e}}
\newcommand {\dd}{\mbox{d}}
\newcommand {\dlangle}{\langle \! \langle}
\newcommand {\drangle}{\rangle \! \rangle}

\newcommand {\defeq}{\stackrel{\rm def}{=}}
\newcommand {\lead}{\stackrel{\rm leading }{\simeq}}

\def\pa{\partial}
\renewcommand{\theequation}{\thesection.\arabic{equation}}

\begin{document}
\setlength{\oddsidemargin}{0cm}
\setlength{\baselineskip}{7mm}

\begin{titlepage}
 \renewcommand{\thefootnote}{\fnsymbol{footnote}}
\begin{normalsize}
\begin{flushright}
\begin{tabular}{l}
UT-Komaba 98-12\\
DPNU-98-22\\
OU-HET 298\\
November 1998
\end{tabular}
\end{flushright}
  \end{normalsize}

~~\\

\vspace*{0cm}
    \begin{Large}
       \begin{center}
         {Dynamical Aspects of
          Large $N$ Reduced Models}      \\
       \end{center}
    \end{Large}
\vspace{1cm}

\begin{center}
           Tomohiro H{\sc otta}$^{1)}$\footnote
            {
e-mail address : 
hotta@hep1.c.u-tokyo.ac.jp, JSPS Research Fellow.},
           Jun N{\sc ishimura}$^{2)}$\footnote
           {
e-mail address : nisimura@eken.phys.nagoya-u.ac.jp,
nisimura@alf.nbi.dk}\footnote{
Present address : Niels Bohr Institute, Blegdamsvej 17, DK-2100, 
Copenhagen \O, Denmark}
           {\sc and}
           Asato T{\sc suchiya}$^{3)}$\footnote
           {
e-mail address : tsuchiya@funpth.phys.sci.osaka-u.ac.jp}\\
      \vspace{1cm}
        $^{1)}$ {\it Institute of Physics, University of Tokyo,}\\
                 {\it Komaba, Meguro-ku, Tokyo 153-8902, Japan}\\

        $^{2)}$ {\it Department of Physics, Nagoya University,}\\
               {\it Chikusa-ku, Nagoya 464-8602, Japan} \\
        $^{3)}$ {\it Department of Physics, Graduate School of  
Science,}\\
               {\it Osaka University, Toyonaka, Osaka 560-0043, Japan}\\
\end{center}

\hspace{5cm}

\begin{abstract}
\noindent
We study the large $N$ reduced model of $D$-dimensional Yang-Mills
theory with special attention to dynamical aspects related to the
eigenvalues of the $N \times N $ matrices, which correspond to the
space-time coordinates in the IIB matrix model.  We first put an upper
bound on the extent of space time by perturbative arguments.  We
perform a Monte Carlo simulation and show that the upper bound is
actually saturated.  The relation of our result to the SSB of the
U(1)$^D$ symmetry in the Eguchi-Kawai model is clarified.  We define a
quantity which represents the uncertainty of the space-time
coordinates and show that it is of the same order as the extent of
space time, which means that a classical space-time picture is
maximally broken.  We develop a $1/D$ expansion, which enables us to
calculate correlation functions of the model analytically.  
%Above all,
%we prove that the large $N$ factorization holds to all orders of the
%$1/D$ expansion.  
The absence of an SSB of the Lorentz invariance is
shown by the Monte Carlo simulation as well as by the $1/D$ expansion.
%as a direct consequence of the large $N$ factorization.
\end{abstract}
\vfill
\end{titlepage}
\vfil\eject

\setcounter{footnote}{0}

\section{Introduction}
\setcounter{equation}{0}
\renewcommand{\thefootnote}{\arabic{footnote}} Large $N$ reduced
models, which are the zero-volume limit of gauge theories, have been
developed for studying the large $N$ limit of gauge theories
\cite{EK,BHN,Parisi,GK,DW}.  Recently the (partially) reduced models have been
revived in the context of nonperturbative formulations of string
theory \cite{BFSS,IKKT}.  The IIB matrix model \cite{IKKT,FKKT,AIKKT},
which is the reduced model of ten-dimensional supersymmetric
Yang-Mills theory, is expected to be a nonperturbative formulation of
superstring theory.  It is hoped that all the fundamental issues
including the origin of the space-time dimensionality can be
understood by studying the dynamics of this model.

Aiming at such an end, we study the large $N$ reduced model of
Yang-Mills theory, which we call as the ``bosonic model'', 
since it is nothing but
the bosonic part of the IIB matrix model.  The action of the bosonic
model is given by
\beq
S= -\frac{1}{4 g^2} \tr ( [A_{\mu},A_{\nu}]^2),
\label{bosonicS}
\eeq
where $A_{\mu}$ $(\mu=1,\cdots,D)$ are $N \times N$ traceless
hermitian matrices.
The partition function is given by
\beq
Z = \int \dd A ~  \ee^{-S},
\label{partitionfunction}
\eeq
where the measure $\dd A$ is defined by
\beq
\dd A =\prod_{\mu=1}^{D}
\left[ \prod_{i=1}^{N} \dd(A_{\mu})_{ii}  
\delta \left(\sum_{i=1}^{N}(A_{\mu})_{ii}\right)
\prod_{i < j}  \dd \mbox{Re}(A_{\mu})_{ij} 
\dd \mbox{Im} (A_{\mu})_{ij} \right].
\label{eq:defmeasure}
\eeq

In Ref. \cite{KS}, 
the value of the partition function of this model 
has been numerically evaluated
and found to be finite\footnote{We will give an intuitive understanding
for their conclusion in Section \ref{upperbound}.} 
for $N \ge 4$ when $D=3$,
for $N \ge 3$ when $D=4$, and
for $N \ge 2$ when $D\ge 5$,
which means that in the large $N$ limit,
the model is well-defined for $D\ge 3$ without any regularizations.
%It is known that perturbation theory gives an attractive potential
%between all the pairs of the eigenvalues of $A_\mu$ when $D > 2$.  Due
%to this, the integral $\int \dd A$ is convergent without any cutoff
%for $D > 2$ when $N$ is sufficiently large\footnote{See Section
%\ref{upperbound}.}.  
%For small $N$ the
%  model can be ill defined for $D=3$ and $4$, as has been found
%  numerically in Ref. \cite{KS}.  We derive analytically the precise
%  condition in which the model is well defined in Section
%  \ref{upperbound}.}.  
This property, together with the fact that the
action is homogeneous with respect to $A_{\mu}$, enables us to absorb
the coupling constant $g$ by rescaling $A_{\mu}'= \frac{1}{\sqrt{g}}
A_{\mu}$.  Hence $g$ is nothing but a scale parameter like the lattice
spacing in lattice gauge theories, and the dependence of expectation
values on $g$ is determined on dimensional grounds, which is also the
case for the IIB matrix model \cite{AIKKT}.  This is in striking
contrast to the ordinary Yang-Mills theory before being reduced, where
the parameter $g$ is the coupling constant, which is independent of
the scale parameter.

We investigate dynamical aspects of the model related to the
eigenvalues of $A_{\mu}$, which correspond to the space-time
coordinates in the IIB matrix model.  One of the most important
quantities is the extent of space time defined by $R=\sqrt{\langle
  \frac{1}{N} \tr (A^{2}) \rangle}$.  Since $R$ should be proportional
to $\sqrt{g}$ on dimensional grounds, we parameterize its large $N$
behavior as $R \sim \sqrt{g} N^{\omega}$.  In the IIB matrix model,
the value of $\omega$ plays an important role when we deduce the
space-time dimension, which is to be determined dynamically within the
model.
% \cite{NT}.  
%It is also important when we consider whether a
%smooth fish-net picture of string worldsheets naturally arises in
%reduced models, which is crucial to the interpretation of these models
%as string theories \cite{NT}.  
We determine the value of $\omega$ for
the bosonic model.  We first obtain an upper bound on $\omega$, namely
$\omega \leq \frac{1}{4}$ by perturbative arguments.  The first
argument is based on a perturbative calculation of the effective action
for the eigenvalues of $A_\mu$.  The second one is based on
exact relations among correlation functions derived as
Schwinger-Dyson equations.  We evaluate the correlation functions 
as a perturbative expansion and 
estimate the large $N$ behavior of each term,
from which we extract an upper bound on the extent of space time 
by requiring that the
Schwinger-Dyson equations should be satisfied if we take all the
terms of the perturbative expansion into account.  
We then perform a Monte Carlo
simulation and find that $\omega=\frac{1}{4}$, which means that the
upper bound given through perturbative arguments is actually
saturated.  This means that 
the perturbative estimation of the large
$N$ behavior of correlation functions
is valid, 
which is also confirmed directly by the Monte Carlo simulation.
We further clarify the relation of our result
concerning the extent of space time to the SSB of the U(1)$^D$
symmetry in the Eguchi-Kawai model.

The space time which is generated dynamically in the IIB matrix model
is not classical, since $A_{\mu}$'s which are given dynamically are
non-commutative generically.  To what extent a classical space-time
picture is broken is therefore an important issue to address.  For
this purpose, we define a quantity which represents the uncertainty of
the space-time coordinates.  We determine its large $N$ behavior for
the bosonic model and show that it is of the same order as the extent
of space time, which means that the classical space-time picture is
maximally broken in the bosonic model.

In the IIB matrix model, the (ten-dimensional) Lorentz
invariance\footnote{When we define the IIB matrix model
  nonperturbatively, we need to make the Wick rotation and define the
  model with Euclidean signature. This is also the case for the
  bosonic model. Therefore, by Lorentz invariance, we actually mean
  the rotational invariance.} should be broken spontaneously if the
space time generated dynamically is to be four-dimensional.  We define
an order parameter for the SSB of the Lorentz invariance.  We show,
for the bosonic model, that the order parameter vanishes in the large
$N$ limit, and hence the Lorentz invariance is not spontaneously
broken.  
%We point out that this result can be viewed as a direct
%consequence of the large $N$ factorization.

As a method complementary to the Monte Carlo simulation, we develop
a $1/D$ expansion, which enables us to calculate correlation
functions of the model analytically.  To all orders of the $1/D$
expansion, we determine the large $N$ behavior of correlation
functions, which agrees with the one obtained by the perturbation
theory and the Monte Carlo simulation.  We also prove that the large
$N$ factorization property holds for the correlation functions.  In
order to confirm the validity of the $1/D$ expansion for studying the
large $N$ limit of the model for finite $D$, we perform explicit
calculations of correlation functions up to the next-leading terms in
$1/D$ and compare the results with the Monte Carlo data.  We conclude
that there is no phase transition at some finite $D$ and that the
$1/D$ expansion is valid for any $D\ge 3$.

This paper is organized as follows. In Section \ref{extentsp}, we
focus on the extent of space time.  We first put an upper bound on
$\omega$, namely $\omega \leq \frac{1}{4}$ by perturbative arguments.
We then show that the upper bound is actually saturated by a Monte
Carlo simulation.  We further clarify the relation of our result to
the SSB of the U(1)$^D$ symmetry in the Eguchi-Kawai model.  In
Section \ref{bd_classics}, we show that a classical space-time
picture is maximally broken in the bosonic model.  In Section
\ref{mftheory}, we develop a $1/D$ expansion, with which we 
determine the large $N$ behavior of correlation functions.
%above all, the large $N$ factorization.  
In Section \ref{lorentzSSB},
the absence of an SSB of the Lorentz invariance is shown by the Monte Carlo
simulation as well as by the $1/D$ expansion.
%as a direct
%consequence of the large $N$ factorization.  
Section \ref{summary} is
devoted to summary and discussion.

\vspace*{1cm}

\section{The extent of space time}
\setcounter{equation}{0}
\label{extentsp}

\subsection{Perturbative arguments}

\subsubsection{Upper bound from the effective action}
\label{upperbound}
We first start by decomposing
$A_{\mu}$ into its eigenvalues $\lambda_{i \mu}$ 
and the angular part $V_{\mu}$
as \cite{GK} 
\beq
A_{\mu}=V_{\mu} \Lambda_{\mu} V_{\mu}^{\dagger}, 
\label{decomp}
\eeq
where $\Lambda_{\mu}=\mbox{diag} (\lambda_{\mu 1},\cdots,\lambda_{\mu N})$ 
and
$V_{\mu}$ is a unitary matrix.
Since $A_\mu$ is traceless, we have $\sum_{i} \lambda_{\mu i} = 0$.
%As we will see shortly, when $\lambda_i$'s are widely separated, 
%$a_\mu$ are suppressed and $A_\mu$ 
The effective action $W(\lambda)$ for the eigenvalues can be defined by
\beq
\ee ^{-W(\lambda)} = 
%\frac{1}{Z} 
\int \dd A \int \dd V \left[ \prod_{\mu} 
\left\{ \prod_{i > j} (\lambda_{\mu i} -\lambda_{\mu j})^2 \right\}
\delta (A_{\mu}-V_{\mu} \Lambda_{\mu} V_{\mu}^{\dagger}) 
\right]
~ \ee^{-S} .
\label{effaction}
\eeq
%The partition function (\ref{partitionfunction}) is given by
%\beq
%Z = \int \dd x ~ \ee^{-W(x)}.
%\eeq
The VEV of an operator ${\cal O}$ can be written as
\beq
\langle {\cal O} \rangle 
= \frac{ \int \dd \lambda \; {\cal O}(\lambda) \; \ee^{-W(\lambda)}}
{\int \dd \lambda  \; \ee^{-W(\lambda)}}  ,
\eeq
where ${\cal O}(\lambda)$ is defined by
\beq
{\cal O}(\lambda)
= \frac{ \int \dd A \int \dd V \left[
\prod_{\mu} 
\left\{ \prod_{i > j}
 (\lambda_{\mu i} -\lambda_{\mu j})^2 \right\}
\delta (A_{\mu}-V_{\mu} \Lambda_{\mu} V_{\mu}^{\dagger}) \right]
 ~ {\cal O}~\ee^{-S}}{\int \dd A \int  
\dd V \left[
\prod_{\mu} \left\{ \prod_{i > j} 
(\lambda_{\mu i} -\lambda_{\mu j})^2 \right\}
\delta (A_{\mu}-V_{\mu} \Lambda_{\mu} V_{\mu}^{\dagger}) \right]
~ \ee^{-S}} .
%{\int \dd a  \; \ee^{-S}}  .
\label{Ox}
\eeq

We can perform a perturbative expansion
with respect to $g^2$ regarding $\lambda_{\mu i}$ as external fields. 
The perturbation theory is valid when $\lambda_i$'s are widely separated,
as we will see shortly.

We first integrate out $V_\mu$ perturbatively around
the unit matrix\footnote{At first sight, one might think that
$V_\mu$ should be expanded not only around 
the unit matrix but also around $E^{(\sigma)}$ whose
matrix elements are given as 
$(E^{(\sigma)})_{ij} = \delta_{\sigma(i)j}$, where
$\sigma$ is an element of $N$-th order symmetric group.
However, one can easily see that
this does not affect the final results for (\ref{effaction})
or (\ref{Ox}).}
either in (\ref{effaction}) or in (\ref{Ox}).
We find that the van der Monde determinants are cancelled out
and obtain the following constraints \cite{GK}.
\beq
d_{\mu i}=\lambda_{\mu i} -\sum_{j \neq i}
\frac{a_{\mu ij} a_{\mu ji}}{\lambda_{\mu i}-\lambda_{\mu j}}
+O(a^3),
\label{constraint}
\eeq
where $d_{\mu i}$ and $a_{\mu ij}$ are the diagonal parts
and the off-diagonal parts of $A_{\mu}$ respectively.

Since the quadratic term in $a$ of the action is
\beq
S_2 = - \frac{1}{2 g^2} 
\{  \tr ([\lambda_\mu,a_\nu]^2) - \tr ([\lambda_\mu,a_\mu]^2) \} ,
\eeq
we have to fix the gauge.
Here we put the following gauge fixing term and the corresponding
Faddeev-Popov ghost term.
\beqa
S_{g.f.} &=& -\frac{1}{2 g^2} \tr ([\lambda_\mu , a_\mu]^2) \\
S_{gh} &=& - \frac{1}{g^2} \tr ([\lambda_\mu , b][d_\mu + a_\mu , c])
\eeqa
The total action $S' = S + S_{g.f.}+S_{gh}$ can be 
written as follows.
\beqa
S'&=&S_{2}'+S_{int}, 
\\
S_{2}' &=& \frac{1}{g^2} \sum_{i \neq j} \left( \frac{1}{2}
                       (\lambda_{i}-\lambda_{j})^2 a_{\mu ji} a_{\mu ij}
       +b_{ji} (\lambda_i-\lambda_j)^2 c_{ij} \right), 
\label{quadratic}
\\
S_{int}&=&-\frac{1}{g^2} \left[ \tr
\left( [\lambda_{\mu},a_{\nu}][a_{\mu},a_{\nu}]
            +\frac{1}{4}[a_{\mu},a_{\nu}]^2
              -b[\lambda_{\mu},[a_{\mu},c]] \right) \right. \n
&& +\sum_{i \neq j,i \neq k}
\left\{ \sum_{\mu,\nu} 
\left( 2\frac{\lambda_{\mu i}-\lambda_{\mu j}}
{\lambda_{\mu i}-\lambda_{\mu k}}
a_{\mu ik} a_{\mu ki} a_{\nu ij} a_{\nu ji} \right.\right. \n
&& \left. 
-\frac{\lambda_{\nu i}-\lambda_{\nu j}}{\lambda_{\mu i}-\lambda_{\mu k}}
a_{\mu ik} a_{\mu ki} ( a_{\mu ij} a_{\nu ji}
+ a_{\mu ji} a_{\nu ij} ) \right) \n
&& \left.\left.
+\sum_{\mu} \frac{\lambda_{\mu i}-\lambda_{\mu j}}
{\lambda_{\mu i}-\lambda_{\mu k}}
a_{\mu ik} a_{\mu ki} (b_{ij} c_{ji} +  b_{ji} c_{ij}) \right\}
+O(a^5) \right]
\eeqa       
From (\ref{quadratic}), we can read off the propagators as
\beqa
\dlangle  a_{\mu ij} a_{\nu kl} \drangle
&=& g^2 \delta_{\mu\nu} 
\delta_{il} \delta_{jk} \frac{1}{(\lambda_i-\lambda_j)^2}, \n
\dlangle c_{ij} b_{kl} \drangle 
&=& g^2 \delta_{il} \delta_{jk} \frac{1}{(\lambda_i-\lambda_j)^2},
\eeqa
where the symbol $\dlangle \cdots \drangle$
is defined by
\beq
\dlangle {\cal O}\drangle =
\frac{\int \dd a  \dd b \dd c ~ {\cal O} ~ \ee ^{-S_2 '}}
{\int \dd a \dd b \dd c ~  \ee ^{-S_2 '}}.
\eeq
The interaction vertices can be read off from $S_{int}$.

The one-loop effective action $W_{1}(\lambda)$ for 
$\lambda_{\mu i}$ is given by
\beq
W_{1}(\lambda)= (D-2) \sum_{i < j} \log (\lambda_i-\lambda_j)^2,
\label{1loopeffectiveaction}
\eeq
which is an $O(N^2)$ quantity 
and gives a logarithmic attractive potential\footnote{As we 
review in Section \ref{relation_to_EK}, 
the bosonic model is actually
equivalent to the weak coupling limit of the Eguchi-Kawai model.
In this correspondence, the logarithmic attractive potential obtained
here for the former model 
is essentially the one obtained \cite{BHN} for 
the latter model.} between all the pairs of $\lambda_{i}$.
If the $\lambda_{i}$'s are sufficiently separated from one another,
the one-loop effective action (\ref{1loopeffectiveaction})
dominates and higher loop corrections can be neglected.
%In this situation, the off-diagonal part $a_\mu$ is suppressed
%by the Gaussian as in (\ref{quadratic}), and the diagonal elements
%$x_{\mu i}$ can be considered as the eigenvalues of the $A_\mu$,
%which represents the space-time coordinates in the IIB matrix model.

We can therefore examine 
the convergence of the integration over $A_\mu$ in
(\ref{partitionfunction})
in the infrared region, where the eigenvalues are far apart,
by using the one-loop effective action
(\ref{1loopeffectiveaction}).
The power of $\lambda$ coming from the one-loop effective action 
is $-(D-2)N(N-1)$,
while the one coming from the measure is $D(N-1)$.
If and only if the sum of the powers is strictly negative,
the integration over $\lambda_{\mu i}$ converges 
in the infrared region.
Thus, we obtain the condition for the infrared convergence as
\beq
N > \frac{D}{D-2}.
\label{condition}
\eeq

Let us next discuss the behavior
of the model in the ultraviolet region,
where some pairs of the eigenvalues are close to each other.
Looking at 
the one-loop effective action (\ref{1loopeffectiveaction}),
one might argue that there might be an ultraviolet divergence.
The two-loop effective action (\ref{twoloop}), 
which we calculate later,
actually shows even a severer divergence for coinciding eigenvalues,
and the higher loop one considers, the severer divergence one encounters.
However, this is simply due to the fact that
the perturbative expansion is no more valid
in the ultraviolet region, and does not imply a real divergence
of the model.
Rather, one should look at the partition function before
performing the perturbative expansion:
%effective action (\ref{effaction}):
\beq
Z =
%\ee ^{-W(\lambda)} = 
%\frac{1}{Z} 
\int \dd \Lambda \int \dd V \left[ \prod_{\mu} 
\left\{ \prod_{i > j} (\lambda_{\mu i} -\lambda_{\mu j})^2 \right\}
\right]
~ \ee^{-S[V_{\mu} \Lambda_{\mu} V_{\mu}^{\dagger}]} ,
\eeq
where one sees no source of divergence for coinciding eigenvalues.

Therefore the model is well defined if and only if the condition
(\ref{condition}) is satisfied. This is in agreement with the results
obtained by a numerical evaluation of
the value of the partition function \cite{KS}.
As far as the large $N$ limit is concerned,
the bosonic model is well defined for any $D \ge 3$.

\begin{figure}[htbp]
  \begin{center}
    \includegraphics[height=5cm]{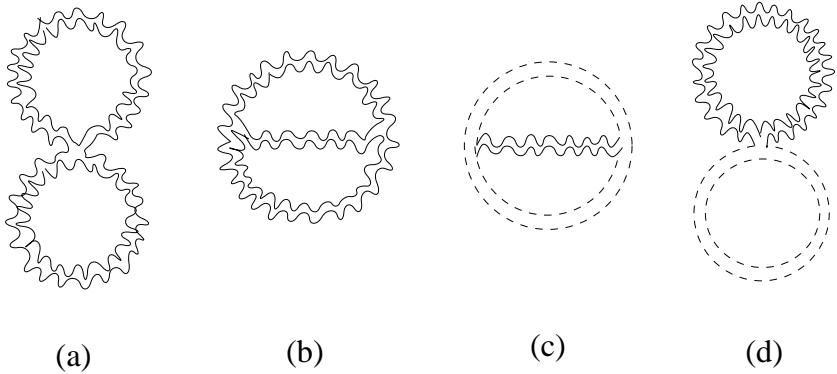}
    \caption{The diagrams we have to consider
to obtain the two-loop corrections to the effective action 
for $\lambda_{\mu i}$.
The wavy line corresponds to $a_{\mu ij}$ and
the dashed line corresponds to the $b$,$c$-ghosts.}
    \label{fig:2loop}
  \end{center}
\end{figure}

As we have seen above, if $\lambda_{i}$'s are widely
separated, the $\lambda_{i}$'s attract one another due to 
the logarithmic attractive potential induced by the one-loop
calculation, and the space time shrinks until the dominance of the
one-loop effective action no more holds.
This means that we can put an upper bound on
the extent of the distribution of $\lambda_i$ by considering 
higher loop corrections to the effective action.

Let us calculate the two-loop corrections.
The diagrams we have to evaluate are shown in Fig. \ref{fig:2loop}.
The diagram (a) is evaluated as follows.
\beqa
(\mbox{a})&=&\frac{1}{g^2} \left\langle \!\! \left\langle
\frac{1}{2} \sum_{ijkl} 
(a_{\mu ij} a_{\nu jk} a_{\mu kl} a_{\nu li}
-a_{\mu ij} a_{\mu jk} a_{\nu kl} a_{\nu li}) \right.\right.\n
&&+\sum_{i \neq j,i \neq k}
\sum_{\mu,\nu} 
\left(
2\frac{\lambda_{\mu i}-\lambda_{\mu j}}{\lambda_{\mu i}-\lambda_{\mu k}}
a_{\mu ik} a_{\mu ki} a_{\nu ij} a_{\nu ji} \right.\n
&& \left.\left.\left.-\frac{\lambda_{\nu i}-\lambda_{\nu j}}
{\lambda_{\mu i}-\lambda_{\mu k}}
a_{\mu ik} a_{\mu ki} a_{\mu ij} a_{\nu ji}
-\frac{\lambda_{\nu i}-\lambda_{\nu j}}{\lambda_{\mu i}-\lambda_{\mu k}}
a_{\mu ik} a_{\mu ki} a_{\mu ji} a_{\nu ij} \right) 
\right\rangle  \!\! \right\rangle \n
&=&  \frac{1}{g^2} 
\left[
\sum_{ijkl} 
\frac{1}{2}  (\dlangle a_{\mu ij} a_{\nu jk} \drangle
   \dlangle  a_{\mu kl} a_{\nu li} \drangle
   - 
\dlangle  a_{\mu ij} a_{\mu jk} \drangle
   \dlangle a_{\nu kl} a_{\nu li} \drangle
 ) \right.\n
&& 
+\sum_{ijk}
\sum_{\mu,\nu} \left\{
2\frac{\lambda_{\mu i}-\lambda_{\mu j}}{\lambda_{\mu i}-\lambda_{\mu k}}
(\dlangle a_{\mu ik} a_{\mu ki} \drangle
\dlangle a_{\nu ij} a_{\nu ji} \drangle
+\dlangle a_{\mu ik}a_{\nu ji} \drangle
\dlangle  a_{\mu ki} a_{\nu ij} \drangle ) \right. \n
&&
-\frac{\lambda_{\nu i}-\lambda_{\nu j}}{\lambda_{\mu i}-\lambda_{\mu k}}
(\dlangle a_{\mu ik} a_{\mu ki} \drangle
\dlangle a_{\mu ij} a_{\nu ji} \drangle
+\dlangle a_{\mu ik}a_{\nu ji} \drangle
\dlangle  a_{\mu ki} a_{\mu ij} \drangle ) \n
&& \left.\left.
- \frac{\lambda_{\nu i}-\lambda_{\nu j}}{\lambda_{\mu i}-\lambda_{\mu k}}
(\dlangle a_{\mu ik} a_{\mu ki} \drangle
\dlangle a_{\mu ji} a_{\nu ij} \drangle
+ \dlangle a_{\mu ik}a_{\mu ji} \drangle
\dlangle  a_{\mu ki} a_{\nu ij} \drangle ) 
\right\} \right] \n
&=&  - \frac{g^2}{2} D(D-1)  \sum_{i\ne j, i\ne k}
 \frac{1}{(\lambda_{i}-\lambda_{j})^2 (\lambda_{i}-\lambda_{k})^2} \n
&&+2 g^2 (D-1) \sum_{i \neq j,i \neq k} 
\sum_{\mu} 
\frac{\lambda_{\mu i}-\lambda_{\mu j}}{\lambda_{\mu i}-\lambda_{\mu k}}
\frac{1}{(\lambda_{i}-\lambda_{j})^2 (\lambda_{i}-\lambda_{k})^2}.
\eeqa
Defining
\beqa
I_1 &=& \sum_{i\ne j, j\ne k,k\ne i}
 \frac{1}{(\lambda_{i}-\lambda_{j})^2 (\lambda_{i}-\lambda_{k})^2},  
\label{exprI12}
\\
I_2 &=& \sum_{i\ne j}
 \frac{1}{(\lambda_{i}-\lambda_{j})^4},
\label{exprI122}
\\
%I_3 &=& \sum_{\mu,\nu} \sum_{i \neq j,i \neq k, j \neq k}
%\frac{x_{\nu i}-x_{\nu j}}{x_{\mu i}-x_{\mu k}}
%\frac{1}{(x_i-x_j)^2 (x_i-x_k)^2}
%\label{exprI3}
%\\
I_3 &=& \sum_{\mu} \sum_{i \neq j,i \neq k, j \neq k}
\frac{\lambda_{\mu i}-\lambda_{\mu j}}{\lambda_{\mu i}-\lambda_{\mu k}}
\frac{1}{(\lambda_i-\lambda_j)^2 (\lambda_i-\lambda_k)^2}  ,
\label{exprI4}
%\\
%I_5 &=& \sum_{\mu,\nu} \sum_{i \neq j}
%\frac{x_{\nu i}-x_{\nu j}}{x_{\mu i}-x_{\mu j}}
%\frac{1}{(x_i-x_j)^4}
%\label{exprI5}
\eeqa
we can write the above result as
\beq
(\mbox{a}) = g^2 \left\{- \frac{1}{2} D(D-1) I_1 + \frac{3}{2} D(D-1) I_2
+2(D-1) I_3 \right\}.
\eeq
The diagrams (b), (c)  and (d) can be calculated in the same way.
The results are
\beqa
(\mbox{b}) &=& 
\frac{3}{2} 
g^2 (D-1) I_1 , \n
(\mbox{c}) &=& -\frac{1}{2} g^2 I_1 ,\n
(\mbox{d}) &=& -2 g^2 (D I_2 + I_3).
\eeqa
Summing up all these results, we obtain the two-loop effective action as
\beq
W_{2}(\lambda)=
-((\mbox{a})+(\mbox{b})+(\mbox{c})+(\mbox{d}))
=  g^2 \left\{
\frac{1}{2} (D-2)^2 I_1 -
\frac{1}{2} D(3 D -7) I_2 
-2(D-2) I_3  \right\}.
\label{twoloop}
\eeq

In order to estimate the order of magnitude of the two-loop 
effective action, we note that
\beq
R^2 = 
\left\langle \frac{1}{N} \sum_i \lambda_i^2 \right\rangle .
\eeq
Assuming that $\lambda_{i}$'s distribute uniformly
in a $D$-dimensional ball with the radius $R$ 
as in Ref. \cite{AIKKT},
we can estimate the order of magnitude of $I_1$ and $I_2$
as follows.
\beqa
I_1 &\sim&  N^3 
\frac{1}{{R}^{3D}} 
\int \dd ^D x ~ \dd ^D y ~ \dd ^D z ~
\frac{1}{(x-y)^2 (x-z)^2}
\sim O\left(\frac{N^3}{{R}^4}\right). 
\label{exprI1}
\\
I_2 &\sim&  N^2 
\frac{1}{{R}^{2D}} 
\int \dd ^D x ~ \dd ^D y ~
\frac{1}{(x-y)^4}
\sim O\left(\frac{N^2}{{R}^4}\right). 
\label{exprI2}
\eeqa
Similarly, we can estimate the order of magnitude of $I_3$ as 
$O\left(\frac{N^3}{{R}^4}\right)$.
Therefore the order of magnitude of the two-loop corrections is
$O\left(\frac{g^2 N^3}{{R}^4}\right)$. 
By simple power counting,
we can see that the $n$-loop corrections are of the order of
$O\left(N^2 \left(\frac{g^2 N}{{R}^4}\right)^{(n-1)}
\right)$, in general.

Since the one-loop result for the effective potential
can be considered as an $O(N^2)$ quantity,
the higher loop corrections can be neglected
as long as $R > \sqrt{g} N ^{1/4}$.
Thus, we can put an upper bound on $R$ as
$R \lesssim \sqrt{g} N ^{1/4}$.
%Note that (\ref{twoloop}) is positive,
%which means that it gives a repulsive potential, which can indeed 
%compensate the attractive potential induced by the one-loop calculation
%when $\tilde{R} \lesssim \sqrt{g} N^{1/4}$.

%The extent $\tilde{R}$ of the distribution of the $x_i$, 
%which is actually not gauge-invariant,
%is not necessarily the same as the extent of space time
%$R=\sqrt{\langle\frac{1}{N}\tr(A^2)\rangle}$
%defined in a gauge-invariant way.
%In order to obtain an upper bound on $R$
%one can redo the perturbative expansion 
%in a gauge-invariant way by decomposing $A_\mu$ as
%\beq
%A_\mu = V_\mu \Lambda_\mu V_\mu^{\dagger},
%\label{gaugeinvdecomp}
%\eeq
%where $\Lambda_\mu$ are real diagonal matrices
%and $V_\mu$ are unitary matrices,
%and expanding $V_\mu$ around the identity matrix.
%Then one obtains the same upper bound on $R$ as that on $\tilde{R}$,
%{\it i.e.}, $R \lesssim \sqrt{g} N^{1/4}$.

%We keep on using the perturbative expansion 
%with the gauge-noninvariant decomposition (\ref{decomp}),
%since the gauge-invariant decomposition (\ref{gaugeinvdecomp})
%cannot be applied straightforwardly to the fermionic matrices 
%in the supersymmetric case.
%However, we note that
%all the statements in the following two sections 
%holds true with $\tilde{R}$ replaced by $R$,
%as can be confirmed by doing the same thing with
%the gauge-invariant decomposition.
%We also see in Section \ref{monte} that the upper bound on
%$R$ and $\tilde{R}$ is actually saturated, 
%which means that they can be identified as long as the
%order in $N$ is concerned.

\subsubsection{Perturbative estimation of correlation functions}
\label{perturb}

In this section, we explain how to estimate the order in $N$ 
of correlation functions of $A$ through perturbative calculations.
We will see that 
the perturbative estimation is valid
if the upper bound on $R$ given in 
Section \ref{upperbound} is actually saturated.
%in which case $R$ and $\tilde{R}$ is of the same order.

Here we perform the integration over $a_\mu$ in (\ref{Ox})
perturbatively as
\beq
{\cal O}(\lambda)
= \frac{ \int \dd a \dd b \dd c ~ {\cal O}~\ee^{-S'}}
{\int \dd a \dd b \dd c  ~ \ee^{-S'}}  .
\label{Oxpert}
\eeq
An example of the operator ${\cal O}$ we consider
is $\tr (A^2)$.
We calculate ${\cal O}(\lambda)$ as a perturbative expansion 
with respect to $g^2$, which gives a loop-wise expansion.
We then perform the integration over $\lambda$
for the result of each order of the perturbative expansion
of ${\cal O}(\lambda)$ with the Boltzmann factor
$\ee ^{-W(\lambda)}$ defined nonperturbatively by (\ref{effaction})

As an example, let us consider the perturbative estimation of 
$\langle\frac{1}{N}\tr((A^2)^2)\rangle$.
\beqa
\left\langle \frac{1}{N}\tr((A^2)^2) \right\rangle
&=&\left\langle \frac{1}{N} \sum_i (\lambda_i^2)^2 \right\rangle
+\left\langle \frac{1}{N} \sum _{ij} \lambda_i^2 |a_{\mu ij}|^2  
\right\rangle+\cdots \n
&=& \left\langle \frac{1}{N} \sum_i (\lambda_i^2)^2 \right\rangle
+\left\langle \frac{1}{N} \sum_{i\ne j} \frac{g^2 \lambda_i^2}
{(\lambda_i-\lambda_j)^2} 
\right\rangle + \cdots.
\label{pert_trA2}
\eeqa 
The order of magnitude of the first and second terms are 
estimated as $O({R}^4)$ and
$O\left(g^2 N\right)$, respectively.
We can easily see that the $n$-loop corrections 
are of the order of
$O\left(
{R}^4 \left(\frac{g^2 N}{{R}^4}\right)^n
\right)$, in general.
Therefore, if the upper bound on $R$ obtained in
Section \ref{upperbound} is actually saturated, 
{\it i.e.}, $R \sim \sqrt{g} N^{1/4}$,
all the orders of the perturbative expansion give
the same order in $N$ and we obtain 
$\langle\frac{1}{N}\tr((A^2)^2)\rangle \sim g^2 N$.

We can generalize the above argument to 
the correlation functions written as
\beq
\left\langle \frac{1}{N}\tr (\underbrace {A\cdots A}_{p_1} )
\frac{1}{N}\tr (\underbrace {A\cdots A}_{p_2} )
\cdots
\frac{1}{N}\tr (\underbrace {A\cdots A}_{p_Q}) \right\rangle .
\eeq
If the upper bound on $R$ is actually saturated, 
the above quantity can be estimated as $(\sqrt{g} N^{1/4})^P$,
where $P=\sum_{i=1}^{Q}  p_i $.

In Section \ref{monte}, we calculate $R$ through a Monte Carlo
simulation and see that the upper bound on $R$
is saturated and that
the perturbative estimation of the large $N$ behavior 
of correlation functions given above is indeed valid.

\subsubsection{Upper bound from Schwinger-Dyson equations}

In this section, we provide another way
to extract information on the extent of space time $R$.
The strategy is to derive exact relations among correlation functions
of $A$ through Schwinger-Dyson equations and to evaluate
each of the correlation functions 
by the perturbative method described in Section \ref{perturb}.
By estimating the large $N$ behavior of the correlation functions,
we will see that the exact relations can be satisfied only if
$R \lesssim \sqrt{g} N^{1/4}$ holds.

We expand $A_{\mu}$ as
\beq
A_{\mu}=\sum_{a=1}^{N^2-1}A_{\mu}^{~a}~t^{a},\\     
\eeq
where $t^a$ are the generators of SU$(N)$.
We consider the following Schwinger-Dyson equations.
\beqa
0&=& 
\int \dd A  \frac{\pa}{\pa A_{\mu}^{~a}} \left( \tr(t^a A_{\mu})
                      \ee^{-S}\right) ,\\
0&=& 
\int \dd A  \frac{\pa}{\pa A_{\mu}^{~a}} 
                   \left( \tr (t^a A_{\nu}A_{\lambda}A_{\rho})
                                   \ee^{-S}\right)   . 
\eeqa
These equations lead to the following exact relations, respectively: 
\beqa
&(\mbox{i})&~~~~~
- \langle \tr(
[A_{\mu},A_{\nu}]^2 ) \rangle
= D (N^2-1)g^2  , 
\label{trF2SD}
\\
&(\mbox{ii})&~~~~~
-g^2 \frac{1}{D} \left\{ \left(N-\frac{1}{N}\right)
(\delta_{\mu\nu}\delta_{\lambda\rho}
+\delta_{\mu\rho}\delta_{\nu\lambda})-\frac{1}{N}\delta_{\mu\lambda}
\delta_{\nu\rho} \right\}\langle \tr(A^2) \rangle \n
&~~~~~&~~~~~~~~~~
=\langle \tr(A_{\nu}A_{\lambda}A_{\rho}[A_{\tau},[A_{\mu},A_{\tau}]])
\rangle . 
\label{SDeq} 
\eeqa
We estimate the large $N$ behavior of 
each of the correlation functions which appear in
the above equations.

Let us first consider eq. (i). 
We calculate the leading term of
the l.h.s. as follows.
\beqa
 -  \langle \tr[A_{\mu},A_{\nu}]^2 \rangle  
&\lead& - 2 \left\{  \langle 
\tr [\lambda_{\mu},a_{\nu}][\lambda_{\mu},a_{\nu}]  \rangle
- \langle \tr[\lambda_{\mu},a_{\nu}]
[\lambda_{\nu} ,a_{\mu}] \rangle \right\}  \n
&=& 
2  \left\{ \left \langle
\sum_{i \neq j} 
(\lambda_{\mu i}-\lambda_{\mu j})(\lambda_{\mu i}-\lambda_{\mu j}) 
\dlangle a_{\nu ji} a_{\nu ij} \drangle 
\right\rangle \right.
 \n
&~& ~~~~~
- 
\left. \left\langle
\sum_{i \neq j} 
(\lambda_{\mu i}-\lambda_{\mu j})(\lambda_{\nu i}-\lambda_{\nu j}) 
\dlangle  a_{\nu ji}a_{\mu ij} \drangle 
\right\rangle \right\}  \n
&=& 2(D-1) N(N-1) g^2.
\label{leading}
\eeqa
The discrepancy between (\ref{leading}) and
the r.h.s. of (i) is an $O(g^2 N^2)$ quantity.
This should be compensated by higher loop corrections
to the l.h.s. of (i).
The typical next-leading contribution of the l.h.s. of (i) is obtained as 
follows.
\beq
\sim g^4 \sum_{i \ne k , j \ne k} 
\frac{1}{(\lambda_i-\lambda_k)^2 (\lambda_j-\lambda_k)^2},
\label{trF2pert}
\eeq
which is an $O(\frac{g^4 N^3}{R^4})$ quantity.
In general, the $n$-th subleading contribution
is $O(g^2 N^2 (\frac{g^2 N}{R^4})^n)$. 
If $R > \sqrt{g} N ^{1/4}$, all the subleading
terms are suppressed compared with the leading term 
and the discrepancy at the leading order cannot be compensated.
We therefore obtain an upper bound 
$R \lesssim \sqrt{g} N ^{1/4}$,
which is the same as the one obtained in Section \ref{upperbound}.

Let us next consider eq. (ii).
The l.h.s. of (ii) can be evaluated to the leading order 
of the perturbative expansion as follows.
\beq
\mbox{l.h.s. of (ii)}\lead - g^2 \frac{1}{D} \left\{
\left( N-\frac{1}{N} \right) 
(\delta_{\mu\nu} \delta_{\lambda\rho}
+\delta_{\mu\rho}\delta_{\nu\lambda})
-\frac{1}{N}\delta_{\mu\lambda}\delta_{\nu\rho} \right\}
\left\langle \sum_{i} \lambda_{i}^{2} \right\rangle.
\label{lhs}
\eeq
The r.h.s. of (ii) can be evaluated as follows.
\beqa
&~& \langle \tr(A_{\nu}A_{\lambda}A_{\rho}[A_{\tau},[A_{\mu},A_{\tau}]])
            \rangle  \n
&&\lead
\langle \tr(a_{\nu}\lambda_{\lambda}\lambda_{\rho}
[\lambda_{\tau},[a_{\mu},\lambda_{\tau}]])
\rangle 
+ \langle 
\tr (\lambda_{\nu}a_{\lambda}\lambda_{\rho}
[\lambda_{\tau},[a_{\mu},\lambda_{\tau}]]) \rangle
\n
&& ~~~~~+ \langle \tr
( \lambda_{\nu}\lambda_{\lambda}a_{\rho}
[\lambda_{\tau},[a_{\mu},\lambda_{\tau}]] )
\rangle
+
\langle \tr
(\lambda_{\nu}\lambda_{\lambda}
\lambda_{\rho}[a_{\tau},[a_{\mu},\lambda_{\tau}]]) \rangle
\n
&& ~~~~~ +
\langle \tr ( a_{\nu}\lambda_{\lambda}\lambda_{\rho}
[\lambda_{\tau},[\lambda_{\mu},a_{\tau}]])
\rangle 
+ \langle \tr
( \lambda_{\nu}a_{\lambda}\lambda_{\rho}
[\lambda_{\tau},[\lambda_{\mu},a_{\tau}]] )
\rangle \n
&& ~~~~~
+ \langle \tr (\lambda_{\nu}
\lambda_{\lambda}a_{\rho}[\lambda_{\tau},[\lambda_{\mu},a_{\tau}]])
\rangle
+ \langle \tr
( \lambda_{\nu}\lambda_{\lambda}
\lambda_{\rho}[a_{\tau},[\lambda_{\mu},a_{\tau}]]) \rangle \n
&&=-g^2\frac{1}{D} \left\{ 
(N-1)(\delta_{\mu\nu} \delta_{\lambda\rho}
+\delta_{\mu\rho}\delta_{\nu\lambda})
-\delta_{\mu\lambda}\delta_{\nu\rho} \right\}
\left \langle \sum_{i} \lambda_{i}^{2} \right \rangle \n
&&~~~~~ -2(D-2) g^2 \left\langle \sum_{i \neq j} 
\frac{(\lambda_{\mu i}-\lambda_{\mu j})\lambda_{\nu i}
\lambda_{\lambda i}\lambda_{\rho i}}
{(\lambda_i-\lambda_j)^2} \right\rangle .
\label{1stterm}
\eeqa
The discrepancy between (\ref{lhs}) and (\ref{1stterm})
which is dominant in the large $N$ limit
comes from the second term in (\ref{1stterm}),
which is estimated as $O(g^2 N^2 R^2)$.
Sub-leading contributions to both sides of (ii)
can be estimated as 
$O(g^2 N^2 R^2 (\frac{g^2 N}{R^4})^n)$.
Thus we obtain the same upper bound on $R$.

%In the supersymmetric model,
%the Schwinger-Dyson equations turn out to be saturated
%by the leading order contributions
%in the large $N$ limit,
% \cite{NT},
%which leads to a different argument on
%the extent of space time.

\subsubsection{Subtlety for the $D=3$ case}
\label{subtlety}

When we estimated the order in $N$ of the expressions
such as (\ref{exprI12}),(\ref{pert_trA2}) and (\ref{trF2pert}),
we considered only the infrared region, in which 
the $x_i$'s are widely separated from one another.
Actually, we have to be careful about the contributions
from the ultraviolet region, in which
the $\lambda_i$'s are close to one another.
As we will see below, the ultraviolet contributions do not affect
the upper bound on $R$ for $D\ge 4$, but they do for $D=3$.
%In the supersymmetric case,
%the ultraviolet contributions indeed become essential
%to the dynamics.
% \cite{NT}.

Let us denote the minimum distance of the $\lambda_i$'s by $r$.
The ultraviolet contributions can be estimated
for (\ref{exprI12}), the second term of (\ref{pert_trA2}), and
(\ref{trF2pert}) as $\frac{N}{r^4}$, $\frac{g^2 R^2}{r^2}$ 
and $\frac{g^4 N}{r^4}$, respectively.
For higher loop corrections, the above expressions 
should be multiplied by $(\frac{g^2}{r^4})^n$,
which means that 
the perturbative arguments break down for $r < \sqrt{g}$.
On the other hand, if $r \gtrsim \sqrt{g}$, 
the ultraviolet contributions can be safely neglected in the 
perturbative arguments given in the previous sections.
One should also consider mixed contributions such 
as the one obtained from (\ref{exprI12})
when $\lambda_i$ and $\lambda_j$ are near 
but $\lambda_i$ and $\lambda_k$ are far apart,
which can be estimated as $\frac{N^2}{r^2 R^2}$.
Such contributions can also be neglected if $r \gtrsim \sqrt{g}$.

In order to estimate the magnitude of $r$,
we introduce the typical distance between two nearest-neighboring 
$\lambda_i$'s, which is given by\footnote{If 
the Lorentz invariance were spontaneously broken,
the $D$ in the definition of $\ell$
should be replaced by the dimension of the space time
dynamically generated. We will see in Section \ref{lorentzSSB}
that the Lorentz invariance is not spontaneously broken in the bosonic
model.} $\ell \sim R \cdot N^{-1/D}$.
In general, we have $r \lesssim \ell $.
If $\ell \gtrsim \sqrt{g}$, 
the distance of the two nearest $\lambda_i$'s
is controlled by an SU(2) matrix model \cite{AIKKT} and therefore,
$r \gtrsim \sqrt{g}$.
Therefore, in order that the perturbative arguments may be valid,
$\ell \gtrsim \sqrt{g}$, or equivalently, 
$R \gtrsim \sqrt{g} N^{1/D}$ must be satisfied.
Taking this point into account,
the upper bound 
$R \lesssim \sqrt{g} N ^{1/4}$
remains unchanged for $D \ge 4$,
but for $D=3$ case, it should be weakened 
to $R \lesssim \sqrt{g} N ^{1/3}$.

\subsection{Monte Carlo calculation of the extent of 
space time and other correlation functions}
\label{monte}

In this section,
we perform a Monte Carlo simulation of the bosonic model
in order to determine the large $N$ behavior of
correlation functions.
Above all, we determine the large $N$ behavior of
the extent of space time, 
defined by $R = \sqrt{ \langle \frac{1}{N} \tr (A^2) \rangle}$.
The details of the algorithm for our simulation are given
in the Appendix.

The observables we measure are the following.
\beqa
(1)&~~~&
\left \langle \frac{1}{N} \tr (A^2) \right\rangle
\times 
\left( \sqrt{g}N^{1/4} \right)^{-2},
~~~~~~~~~~~~~~~~~~~~~~~~~~~~~~~~~~~~~~~ \n
(2)&~~~&
\left\langle \frac{1}{N} \tr ((A_\mu A_\nu ) ^2 ) 
\right\rangle \times 
\left( \sqrt{g}N^{1/4} \right)^{-4},
~~~~~~~~~~~~~~~~~~~~~~~~~~~~~~~~~~~~~~~ \n
(3)&~~~& 
\left\langle \frac{1}{N} \tr ((A^2)^2) \right\rangle \times 
\left( \sqrt{g}N^{1/4} \right)^{-4},
~~~~~~~~~~~~~~~~~~~~~~~~~~~~~~~~~~~~~~~ \n
(4)&~~~&
\left\langle \frac{1}{N} \tr (F ^2) \right\rangle \times 
\left( \sqrt{g}N^{1/4} \right)^{-4},
~~~~~~~~~~~~~~~~~~~~~~~~~~~~~~~~~~~~~~~
\eeqa
where we define $F_{\mu\nu}= i[A_\mu, A_\nu]$.
Actually $(4)$ can be written as $(4) = 2 \times ((3)-(2))$.
Note that the above quantities are independent of $g$,
since we have normalized them so that
they are dimensionless.

We first note that 
$\left\langle \frac{1}{N} \tr (F ^2) \right\rangle$ 
can be obtained analytically by the following scaling argument.
Rescaling $A_{\mu}$ 
as $A_{\mu} \rightarrow \ee^{\epsilon} A_{\mu}$
in the definition of $Z$ in (\ref{partitionfunction}), 
we obtain
\beqa
Z&=&\ee^{\epsilon D (N^2-1) }  
\int \dd A ~
      \ee^{- \frac{1}{4 g^2} \ees^{4 \epsilon} \trs(F^2) }\n
 &=&Z+\epsilon \left \{ D (N^2-1) Z 
    -\frac{1}{g^2} \int \dd A ~  \tr(F^2) ~
\ee^{-S} \right\} +  O(\epsilon^2).
\eeqa
Thus we obtain
\beq
\left\langle \frac{1}{N} \tr(F^2) \right\rangle
= D N \left(1-\frac{1}{N^2} \right) g^2.
\label{trf2}
\eeq
This agrees with the result (\ref{trF2SD})
obtained through the Schwinger-Dyson equation.

\begin{figure}[htbp]
  \begin{center}
    \includegraphics[height=10cm]{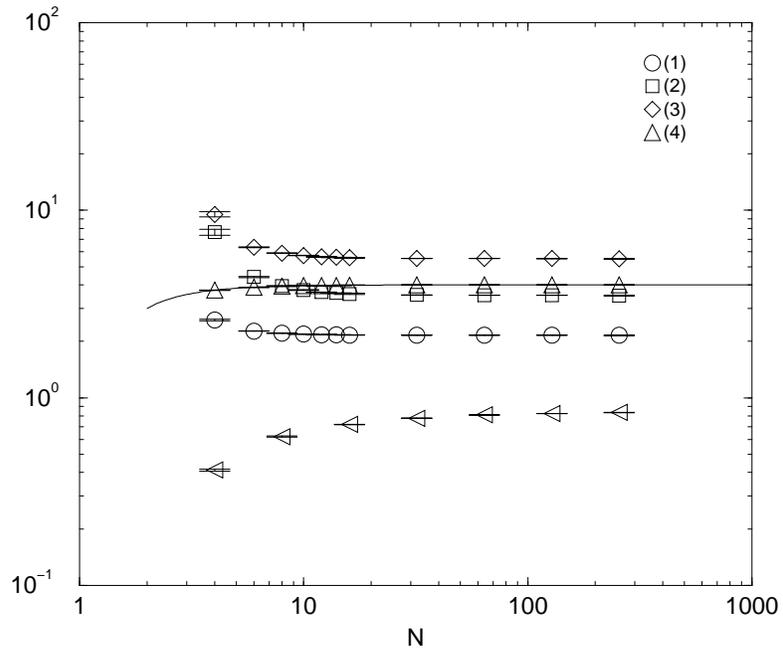}
    \caption{The quantities (1)(circles),(2)(squares),(3)(diamonds)
      and (4)(triangles) are plotted against $N$ for the bosonic model
      with $D=4$.  The solid line represents the exact result
      (\ref{trf2}) for (4).  The tilted triangles are the plot of
      $\left\langle \Delta ^2 \right \rangle / (\sqrt{g}N^{1/4})^2$,
      which is discussed in Section \ref{bd_classics}.}
    \label{fig:D4}
  \end{center}
\end{figure}

\begin{figure}[htbp]
  \begin{center}
    \includegraphics[height=10cm]{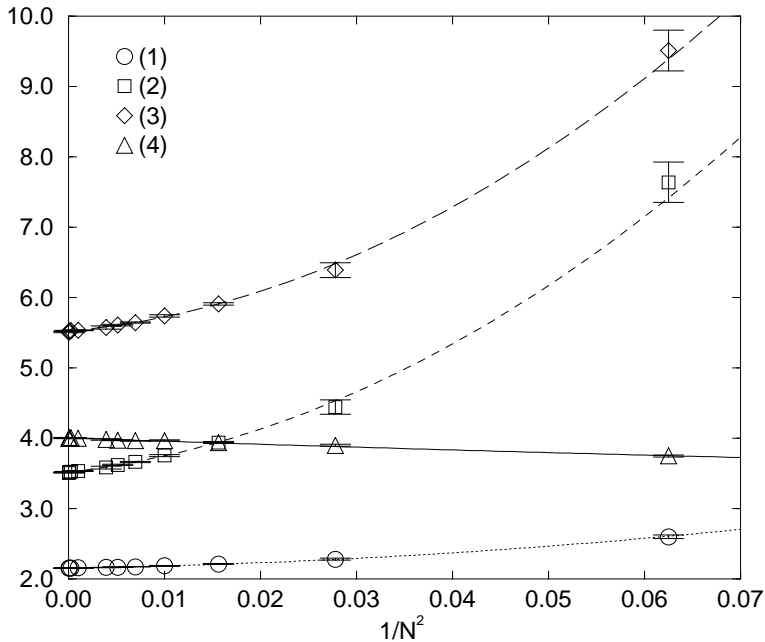}
    \caption{The quantities (1)(circles),(2)(squares),(3)(diamonds)
      and (4)(triangles) are plotted against $1/N^2$ for the bosonic
      model with $D=4$.  The solid line represents the exact result
      (\ref{trf2}) for (4).  The other lines represent the fits to
      $a_0 +\frac{a_1}{N^2}+\frac{a_2}{N^4}$.}
    \label{fig:D4fit}
  \end{center}
\end{figure}

In Fig. \ref{fig:D4}, we show our results of the Monte Carlo
simulation for $D=4$ with $N=4 \sim 256$.  One can see that the
quantities (1)$\sim$(4) are constant for $N \gtrsim 16$.
In order to see the finite $N$ effects, we plot in Fig.
\ref{fig:D4fit} the same data against $1/N^2$.  The data can be fitted
nicely to $a_0+\frac{a_1}{N^2}+\frac{a_2}{N^4}$.  In Section
\ref{largeNfactorization}, we will see that this large $N$ behavior of
the observables (1)$\sim$(4) agrees with the one given by the $1/D$
expansion.

\begin{figure}[htbp]
  \begin{center}
    \includegraphics[height=10cm]{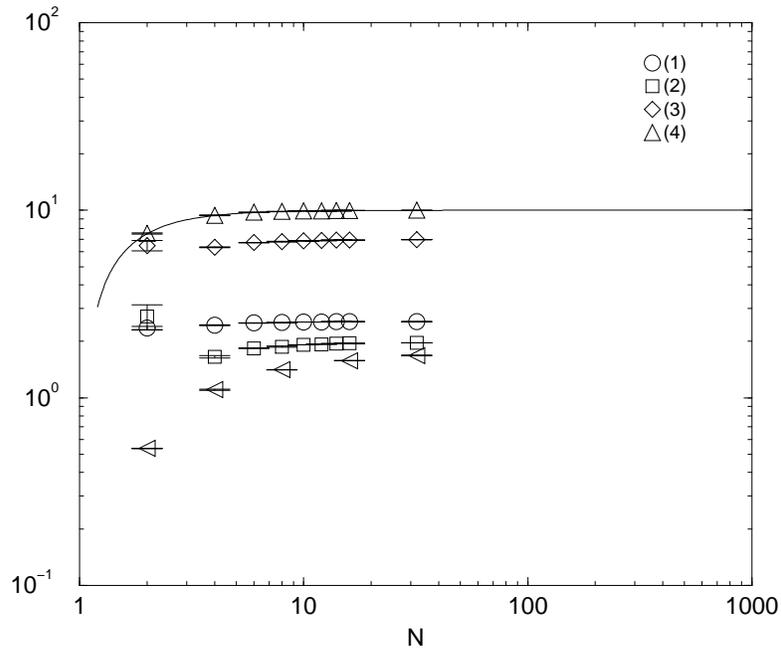}
    \caption{The quantities (1)(circles),(2)(squares),(3)(diamonds)
      and (4)(triangles) are plotted against $N$ for the bosonic model
      with $D=10$.  The solid line represents the exact result
      (\ref{trf2}) for (4).  The tilted triangles are the plot of
      $\left\langle \Delta ^2 \right \rangle / (\sqrt{g}N^{1/4})^2$,
      which is discussed in Section \ref{bd_classics}.}
    \label{fig:D10}
  \end{center}
\end{figure}

In Fig. \ref{fig:D10},
we plot the data for $D=10$ with $N=2 \sim 32$.
Here again we find that 
the quantities (1)$\sim$(4) are constant 
for $N \gtrsim 16$. We have also checked that the
data can be fitted nicely to $a_0+\frac{a_1}{N^2}+\frac{a_2}{N^4}$.

It is rather surprising that the leading large $N$ behavior shows up
at about the same $N$ for $D=4$ and $D=10$.  If the system were a gas
model with $N$ particles in a $D$-dimensional space time, the large
$N$ behavior would be expected naively to show up at such $N$ that
satisfies $N^{1/D} \gg 1$.  In Section \ref{bd_classics}, we define
the uncertainty of the space-time coordinates and show that it is of
the same order as the extent of space time, which means that the
bosonic model indeed can hardly be considered as a gas model with $N$
particles in a $D$-dimensional space time.

From the results of the Monte Carlo simulation, we find that the
extent of space time $R=\sqrt{\langle\frac{1}{N}\tr(A^2)\rangle}$ is
of the order of $O(\sqrt{g}N^{1/4})$, which means that the upper bound
on $R$ given in the previous sections for $D\ge 4$ is actually
saturated.  
%We have also measured $\tilde{R}$
%for $D=4$ and $D=10$ and found that it is of the same order as $R$.
Due to this, the perturbative estimation of the large $N$
behavior of correlation functions given in Section \ref{perturb} with
the input of $R \sim O(\sqrt{g}N^{1/4})$ 
is valid for $D\ge 4$, which is also confirmed directly 
by the Monte Carlo simulation.

\begin{figure}[htbp]
  \begin{center}
    \includegraphics[height=10cm]{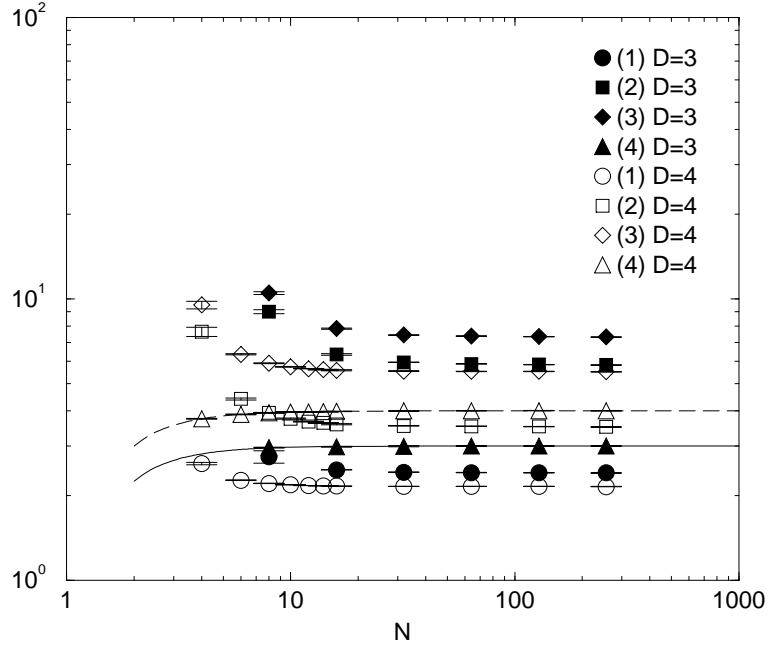}
    \caption{The quantities (1)(circles),(2)(squares),(3)(diamonds)
      and (4)(triangles) are plotted against $N$ for the bosonic model
      with $D=3$.  The results for the $D=4$ are replotted for
      comparison.  The solid line and the dashed line represent the
      exact result (\ref{trf2}) for (4) for $D=3$ and $D=4$,
      respectively.}
    \label{fig:D3a4}
  \end{center}
\end{figure}

We have done a simulation for the $D=3$ case also, since it is rather
an exceptional case from the perturbative point of view, as we
remarked in Section \ref{subtlety}.  The result is shown in Fig.
\ref{fig:D3a4}.  One can see that the result is qualitatively the same
as for the $D=4$ case.  The fact that the upper bound $R \lesssim
\sqrt{g}N^{1/3}$ is not saturated means that the perturbative
arguments completely break down in the $D=3$ case due to the
ultraviolet contributions.  We will see in Section \ref{comparison},
however, that the $1/D$ expansion is valid even for $D=3$, which
explains why the results for $D=3$ are qualitatively the same as those
for $D \ge 4$.

\subsection{Relation to the U(1)$^D$ SSB of the EK model}
\label{relation_to_EK}

In this section, we examine the relation of our result concerning the
extent of space time to the U(1)$^D$ SSB of the $D$-dimensional
Eguchi-Kawai model.

The action of the $D$-dimensional Eguchi-Kawai model \cite{EK} is
given by
\begin{equation}
  \label{red-action}
  S_{EK}=-N\beta \sum_{\mu \ne \nu=1}^{D} 
\tr (U_\mu U_\nu U^{\dag}_\mu U^{\dag}_\nu ).
\end{equation}
This model has a U(1)$^{D}$ symmetry.
\begin{equation}
U_\mu \to \ee^{i \theta_\mu} U_\mu.
\label{usym}
\end{equation}
In Ref. \cite{EK}, it has been shown that if the U(1)$^{D}$ symmetry
is not spontaneously broken, the model is equivalent to an SU($N$)
lattice gauge theory on an infinite lattice in the large $N$ limit,
where the coupling constant $\beta$ in the action (\ref{red-action})
is kept fixed.

It is found in Ref. \cite{BHN} that when $D$ is larger than two, the
U(1)$^{D}$ symmetry is spontaneously broken in the weak coupling
(large $\beta$) region.  This means that we cannot study the continuum
limit of the large $N$ gauge theory by the EK model when $D>2$, since
we have to send $\beta$ to infinity when we take the continuum limit.
Slightly modified versions of the model known as the quenched EK model
\cite{BHN,GK,DW} and the twisted EK model \cite{GO} have been shown to be
equivalent to the SU($N$) lattice gauge theory on the infinite lattice
in the large $N$ limit.

Here we note that the bosonic model is actually equivalent to the weak
coupling limit of the original EK model for $D>2$.  This can be seen
as follows \cite{BHN}.  Since $U_\mu$ is unitary, the eigenvalues of
$U_\mu$ are on a unit circle in the complex plane.  The U(1)$^{D}$
symmetry rotates all the eigenvalues by the same angle.  In the weak
coupling limit, namely when $\beta \rightarrow \infty$, the U(1)$^{D}$
symmetry is spontaneously broken and the eigenvalues collapse to a
point.  Therefore, when $\beta$ is sufficiently large, the dominant
configurations are given by
\beq
U_\mu \sim \ee ^{i \alpha_\mu}\ee ^{i A_\mu},
\label{degenerate}
\eeq
where $\alpha _\mu$ are real numbers defined modulo $2 \pi$ and
$A_\mu$ are $N \times N$ traceless hermitian matrices, whose
eigenvalues are small.  $\alpha_\mu$ take random values due to the
U(1)$^D$ symmetry.  The action for $A_\mu$ is obtained by putting
(\ref{degenerate}) in (\ref{red-action}) and expanding it in terms of
$A_\mu$. We obtain the following.
\begin{equation}
  \label{red-action2}
  S_{EK} \sim -  \frac{1}{2} N \beta \sum_{\mu \ne \nu=1}^{D} 
\tr ([A_\mu , A_\nu ] ^2) + \mbox{(higher order terms in $A$)}.
\end{equation}
In the weak coupling limit, the higher order terms 
can be neglected and
we arrive at the bosonic model (\ref{bosonicS}) 
with $g=\frac{1}{\sqrt{2N\beta}}$.

A quantity which represents the extent of the eigenvalue distribution
in the Eguchi-Kawai model can be defined by \cite{BHN}
\beq
  \label{defP}
P = \frac{1}{2} \left\{ 
1 - \frac{1}{D N^2} \sum _{\mu}
| \tr ~ U_\mu |^2  \right\}  .
\eeq
As can be seen in the above expression, $0 \le P \le \frac{1}{2}$ for
any $U_\mu$.  $P \sim \frac{1}{2}$ when the eigenvalues are uniformly
distributed, while $P \sim 0$ when the eigenvalues collapse.  Thus
$\langle P \rangle$ serves as an order parameter for the SSB of the
U(1)$^D$ symmetry.

Let us consider $\langle P \rangle$ in the weak coupling limit.
Putting (\ref{degenerate}) in (\ref{defP}) and expanding it in terms
of $A_\mu$, we obtain, to the leading order of $A_\mu$,
\beq
  \label{defP2}
P \sim \frac{1}{2DN} \sum_{\mu} \tr (A_\mu ^{~2}) .
\eeq
Therefore $\langle P \rangle \sim \frac{1}{2D} R^2$, 
where $R$ is the extent of the eigenvalue distribution
defined in the bosonic model.
In Section \ref{monte}, we have seen that
$R^2 = C (\sqrt{g} N^{1/4})^2$ in the large $N$ limit.
This means that
%The fact that $R^2 \sim C (\sqrt{g} N^{1/4})^2$ means that
\beq
\langle P \rangle \sim \frac{C}{2\sqrt{2} D} 
\frac{1}{\beta^{1/2}},
\label{predictionP}
\eeq
as $\beta \rightarrow \infty $.
Thus our finding for the bosonic model can be interpreted in
terms of the EK model as
the eigenvalue distribution having a finite extent in the large 
$N$ limit for fixed $\beta$ in the weak coupling region.
We check this explicitly by a Monte Carlo simulation in what follows.

A Monte Carlo simulation of the EK model with $D=4$ and $N=20$
has been performed in Ref. \cite{Okawa},
where the internal energy defined by
\beq
E=\left\langle 
\frac{1}{N} \frac{1}{D(D-1)}
\sum_{\mu\ne\nu} \tr(U_{\mu}U_{\nu}U_{\mu}^{\dagger}
U_{\nu}^{\dagger})
\right\rangle 
\label{predictionE}
\eeq
is measured.
The asymptotic behavior of the quantity in the weak coupling limit 
can be obtained just in the same way as we derived (\ref{predictionP}).
The result to the leading order of $A_\mu$ reads \cite{Okawa}
\beqa
E &\sim& 1+ \frac{1}{2ND(D-1)}
\left\langle
\sum_{\mu\ne\nu} \tr([A_\mu,A_\nu]^2)
\right\rangle  \n
&=& 1- \frac{1}{4(D-1)}\left( 1 - \frac{1}{N^2} \right) 
\frac{1}{\beta},
\label{eq:meanplaquette_asym}
\eeqa
where we used the exact result (\ref{trf2}) in the 
second equality.

\begin{figure}[htbp]
  \begin{center}
    \includegraphics[height=10cm]{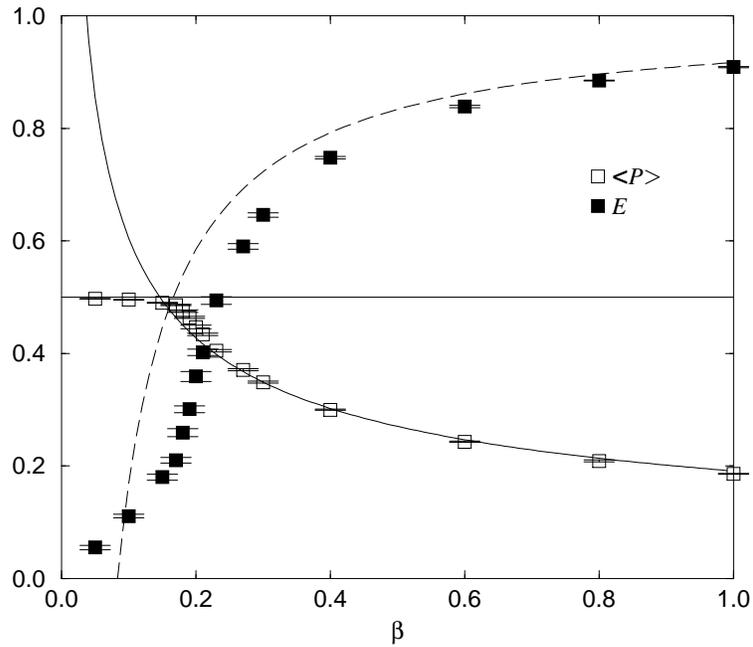}
    \caption{The order parameter $\langle P \rangle$
      for the U(1)$^D$ symmetry (open squares) as well as the internal
      energy $E$ (filled squares) is plotted against $\beta$ for the
      $D=4$ SU(16) Eguchi-Kawai model. The solid line and the dashed
      line represent the asymptotic behavior of $\langle P \rangle$
      and $E$ respectively at large $\beta$ predicted by the bosonic
      model.}
    \label{fig:EK}
  \end{center}
\end{figure}

In Fig. \ref{fig:EK}, we plot the order parameter $\langle P \rangle$
as well as the internal energy $E$ for the SU(16) EK model.  The solid
line and the dashed line represent the asymptotic behavior at large
$\beta$ for $\langle P \rangle$ (\ref{predictionP}) with $C=2.162(5)$
and that for $E$ (\ref{eq:meanplaquette_asym}), respectively,
predicted by the bosonic model.  The data for large $\beta$ fit nicely
to the predictions.  The results for the internal energy are in good
agreement with the results in Ref. \cite{Okawa}, though $N$ differs
from ours slightly.  We have also done simulations with $N=32$ and
$N=64$ for $\beta = 1.0$ and checked that both $E$ and $\langle P
\rangle$ remain the same within error bars.  This large $N$ behavior
is just the one we expect from the large $N$ behavior of the
quantities (1) and (4) in the bosonic model obtained in Section
\ref{monte}.

$\langle P \rangle$ is around 0.5 for $\beta \lesssim 0.17$ and
decreases for $\beta \gtrsim 0.17$, which shows that the U(1)$^D$
symmetry is spontaneously broken at around $\beta = \beta _c \sim
0.17$.  At the critical point $\beta = \beta _c$, the internal energy
is continuous but its first derivative with respect to $\beta$ seems
to diverge, which suggests that the system undergoes a second order
phase transition.

It is intriguing to compare the phase diagram of the EK model with
that of the TEK model \cite{GO,JN}.  Since the U(1)$^D$ symmetry is
not spontaneously broken in the TEK model even in the weak coupling
region, the model is equivalent to the SU($N$) lattice gauge theory on
the infinite lattice in the large $N$ limit throughout the whole
region of $\beta$ \cite{GO}.  The system undergoes a first order phase
transition at $\beta \sim 0.35$.  This phase transition can be
considered as an artifact of the Wilson's plaquette action, and could
be avoided if one wishes, say, by introducing the adjoint term
\cite{JN}.

If the $\beta_c$, at which the SSB of the U(1)$^D$ symmetry occurs in
the EK model, were larger than 0.35, the EK model would also have
undergone a first order phase transition at $\beta \sim 0.35$, but the
fact is that $\beta_c < 0.35$ and the model is not equivalent to the
SU($N$) lattice gauge theory for $\beta > \beta_c$.  Thus the first
order phase transition may well be absent in the EK model.

Before ending this section, let us address the issue \cite{KS} whether
the bosonic model can be used to calculate correlation functions in
the large $N$ Yang-Mills theory.  It is natural to think that the
answer is in the negative considering the fact that the EK model is
not equivalent to the large $N$ lattice gauge theory for $D>2$ due to
the SSB of the U(1)$^D$ symmetry.  Let us see how this can be restated
within the bosonic model.

One way to show the equivalence between the reduced model and the
original model in the large $N$ limit is to confirm it to all orders
of the perturbative expansion \cite{Parisi,GK,DW},
%Indeed the diagrams one
%encounters in the perturbative expansion of the bosonic model
%described in Section \ref{upperbound} have a one-to-one correspondence
%with the ones in the perturbative expansion of the Yang-Mills theory,
where $\lambda_{\mu i}$ should be identified with the loop momenta and the
extent of space time $R$ plays the role of the momentum cutoff.  One
can easily see that the coupling constant $g_{YM}$ of the
corresponding Yang-Mills theory is given by
\beq
g_{YM}^{~2}=\frac{g^2}{R^D},
\label{gFT}
\eeq
which can be also deduced on dimensional grounds.
In order to take a nontrivial large $N$ limit, namely the 
't Hooft limit,
one has to fix $g_{YM}^{~2} N = \lambda_{YM}$ to a constant.
Since $R = C^{1/2} \sqrt{g}N^{1/4}$, where $C$
is a fixed constant determined dynamically, we obtain
\beq
\lambda_{YM}= C^{-\frac{D}{2}}  (g^2 N)^{1-\frac{D}{4}} .
\label{YM}
\eeq
From (\ref{YM}),
we have to fix $g^2 N$ to reproduce the 't Hooft limit.
In this limit, $R$ goes to a constant, which we denote by $\Lambda$.
$\lambda_{YM}$ can then be written as
\beq
\lambda_{YM} = C^{-2} \Lambda ^{4-D},
\eeq
which means that we are not allowed to take the continuum limit
$\Lambda \rightarrow \infty$ as one wishes, but the coupling constant
$\lambda_{YM}$ is doomed to scale canonically with the cutoff
$\Lambda$.  Thus, there is no way to take a nontrivial continuum limit
in the Yang-Mills theory to which the bosonic model is equivalent
perturbatively\footnote{The equivalence does not hold in the strict
sense even perturbatively. See Ref. \cite{GK} for the details.}.  
We therefore conclude that the bosonic model
cannot be used to calculate correlation functions in the large $N$
Yang-Mills theory.

\vspace*{1cm}

\section{Breakdown of the classical space-time picture}
\setcounter{equation}{0}
\label{bd_classics}

As we mentioned in the Introduction, the eigenvalues of $A_{\mu}$
represent the space-time coordinates, when we interpret the IIB matrix
model as a string theory formulated nonperturbatively.  If $A_{\mu}$'s
are mutually commutative, they can be diagonalized simultaneously and
the diagonal elements after diagonalization can be regarded as the
classical space-time coordinates.  This must be violated more or less
by $A_{\mu}$'s generated dynamically in the IIB matrix model.
Therefore it makes sense to ask to what extent the classical
space-time picture is broken.  In order to address this issue, we
define a quantity which represents the uncertainty of the space-time
coordinates.  We determine its large $N$ behavior for the bosonic
model and show that it is of the same order as the extent of space
time, which means that the classical space-time picture is maximally
broken in the bosonic model.

We define such a quantity by considering an analogy to the quantum
mechanics.
We regard the matrices $A_\mu$'s as linear operators which
act on a linear space, which we identify as the space of states
of particles.
We take an orthonormal basis $|e_i \rangle$ ($i=1,2,\cdots,N$) 
of the $N$-dimensional linear space, and
identify the ket $|e_i \rangle$ with the state of the $i$-th particle.
The space-time coordinate of the $i$-th particle can be defined
by $\langle e_i | A_\mu | e_i \rangle $.
The uncertainty $\delta (i)$ of the space-time coordinate 
of the $i$-th particle can be defined by
\beq
\delta (i)^2
= \sum_{\mu} \left\{
\langle e_i | A_\mu ^{~2} | e_i \rangle
- (  \langle e_i | A_\mu | e_i \rangle )^2  \right\}.
\eeq
Note that this quantity is invariant under a Lorentz transformation
and a translation : $A_\mu \rightarrow A_\mu + \alpha_\mu$.
We take an average of $\delta (i)^2$ over all the $N$ particles.
\beq
\overline{\delta ^2} = \frac{1}{N} \sum_{i} \delta(i) ^2.
\eeq
Note that this quantity depends on 
the orthonormal basis $|e_i \rangle$ 
($i=1,2,\cdots,N$) we choose.
We therefore define a quantity $\Delta$
which represents the uncertainty of the space-time 
coordinates\footnote{It might be tempting 
to identify the quantity $\Delta$
with the space-time uncertainty which appears in the space-time
uncertainty principle proposed in Ref. \cite{Yoneya}.
However, it is unclear at present 
whether this identification is correct.} for a given $A_\mu$ by
\beqa
\Delta ^2 &=& \min_{\{|e_i \rangle\} } (\overline{\delta ^2}) \\
         &=& \frac{1}{N} \tr (A_\mu^{~2})
- \max_{U \in \mbox{\scriptsize SU}(N)} \frac{1}{N}
\sum_i \{ (U A_\mu U^\dagger)_{ii}  \} ^2.
\label{maxU}
\eeqa
Note that $\Delta$ is invariant under a gauge transformation:
$A_\mu \rightarrow g A_\mu g^{\dagger}$.
Note also that $\Delta = 0$ if and only if $A_\mu$ are
diagonalizable simultaneously.
The classical space-time picture is good
if $\Delta$ is smaller than the typical distance 
$\ell \sim R \cdot N^{-1/D}$ between two 
nearest-neighboring particles.

%Let us evaluate the quantity $\Delta$ in the bosonic model.
%We first put an upper bound on $\langle  \Delta ^2 \rangle$ 
%using the perturbative argument given in Section \ref{perturb}.
%Noting that
%\beq
%\max_{U \in \mbox{\scriptsize SU}(N)} \frac{1}{N}
%\sum_i
%\{ (U A_\mu U^\dagger)_{ii}  \} ^2
%\ge \frac{1}{N} \sum _i (x_{\mu i})^2,
%\eeq
%we obtain
%\beq
%\langle \Delta ^2 \rangle  \le  
%\left\langle  \frac{1}{N} \tr (A^2) \right\rangle
%- \left\langle
%\frac{1}{N} \sum _i (x_{\mu i})^2 \right\rangle 
%= \left\langle
%\frac{1}{N} \sum _{ij} | a_{\mu i j}|^2 \right\rangle .
%\label{Deltaub}
%\eeq
%As can be seen by the argument below eq. (\ref{pert_trA2}),
%the quantity on the r.h.s. of the above inequality
%is of the order of $O(g N^{\frac{1}{2}})$,
%which is of the same order as $R^2$. 
%Thus we obtain the upper bound
%on $\sqrt{\langle \Delta ^2 \rangle}$ as
%$\sqrt{\langle \Delta ^2 \rangle} \lesssim O(R) $.
%In what follows, 
We calculate 
$\langle \Delta ^2 \rangle$ 
for the bosonic model
by a Monte Carlo simulation.
%and show that the upper bound
%is actually saturated, which 
%means that the classical space-time picture
%is maximally broken in the bosonic model.
Given a configuration $A_\mu$, which is generated by a Monte Carlo
simulation, we evaluate $\Delta ^2 $ numerically by performing the
maximization of $\sum_i\{ (U A_\mu U^\dagger)_{ii} \} ^2$ in
(\ref{maxU}) in the following way.  We first maximize $\sum_i \{ (U
A_\mu U^\dagger)_{ii} \} ^2 $ restricting the $U$ to be in one of the
SU(2) subgroup of the SU($N$).  Redefining the $A_\mu$ by the $U_0
A_\mu U_0^\dagger$, where $U_0$ is the element of the SU(2) subgroup
that gives the maximum, we do the maximization for all the $~_N
\mbox{C}_2$ SU(2) subgroups successively.  We call this as `one sweep'
in our maximization procedure.  For $N \le 32$, the quantity saturates
up to 16 digits within 200 sweeps for $D=4$, and within 500 sweeps for
$D=10$.  For $D=4$, we have done the measurement with larger $N$ as
well.  For $N=256$, we have made 300 sweeps of the maximization, with
which the saturation is achieved up to 3 digits.  The result thus
obtained is plotted in Figs. \ref{fig:D4} and \ref{fig:D10}.  We find
that $\langle \Delta ^2 \rangle$ normalized by $(\sqrt{g}N^{1/4})^2$
tends to a constant for large $N$.  Thus we conclude that
$\sqrt{\langle \Delta ^2 \rangle}$ is of the same order as $R$.

We have seen in Section \ref{monte} that the $N$ at which the leading
large $N$ behavior dominates does not depend much on $D$.  The large
$N$ behavior would show up for $N^{1/D} \gg 1 $ if the system can be
viewed as $N$ classical particles in a $D$-dimensional space time.
Indeed our conclusion in this section means that the bosonic model can
hardly be viewed as consisting of $N$ particles.

\vspace*{1cm}

\section{$1/D$ expansion of the bosonic model}
\label{mftheory}
\setcounter{equation}{0}

\subsection{The formalism}
\label{largeDexp}

In this section, we show that the bosonic model allows a systematic
$1/D$ expansion, which serves as an analytical method complementary to
the Monte Carlo simulation.  To all orders of the $1/D$ expansion, we
determine the large $N$ behavior of correlation functions and prove
the large $N$ factorization property.  We also perform an explicit
calculation of the quantities (1)$\sim$(4) defined in Section
\ref{monte} up to the next-leading order and compare the results with
the Monte Carlo data for various $D$, from which we conclude that the
$1/D$ expansion is valid down to $D=3$.

Here we use again the adjoint notation defined through
\beq
A_{\mu}=\sum_{a=1}^{N^2-1} A_{\mu}^{~ a} ~ t^{a},
\eeq
where $t^a$ are the generators of SU($N$).
Taking the generators $t^a$ to satisfy the orthonormal condition
\beq
\tr (t^a t^b) = \delta _{ab},
\eeq
we have the following relation
\beq
\sum_{a} (t^a)_{ij} (t^a)_{kl} =
 \delta _{il} \delta _{jk} - \frac{1}{N} \delta _{ij} \delta _{kl}.
\label{Trformula}
\eeq

We first rewrite the action (\ref{bosonicS})
for the bosonic model as
\beqa
S &=& - \frac{1}{4 g^2} \tr ( [A_\mu, A_\nu ]^2 )\n
&=& - \frac{1}{4 g^2} \lambda ^{abcd}
A_\mu^{~a}A_\mu^{~b}A_\nu^{~c}A_\nu^{~d},
\label{originalaction}
\eeqa
where we have defined
\beq
\lambda ^{abcd} =  \frac{1}{4} \left\{ \tr ( [t^a,t^c][t^b,t^d])
+ (a\leftrightarrow b)
+ (c\leftrightarrow d)
+ \left( \begin{array}{c}
a\leftrightarrow b  \\
c\leftrightarrow d
\end{array}
\right)  \right\}.
\eeq
Note that the measure (\ref{eq:defmeasure}) can be written
as $\prod_{\mu a} \dd A_\mu^{~a}$ up to an irrelevant constant factor. 
We introduce the auxiliary field $h_{ab}$,
which is a real symmetric tensor, with the following action.
\beqa
S[A,h] & =& \frac{1}{4 g^2} \lambda ^{abcd}
( h_{ab} h_{cd} - A_\mu^{~a}A_\mu^{~b} h_{cd} 
-  h_{ab} A_\nu^{~c}A_\nu^{~d} ) \\
&=& \frac{1}{4 g^2} \lambda ^{abcd}
h_{ab} h_{cd} + \frac{1}{2g} K_{ab} A_\mu^{~a}A_\mu^{~b},
\label{auxiliaryaction}
\eeqa
where we have defined the dimensionless kernel
\beq
 K_{ab} = - \frac{1}{g} \lambda ^{abcd} h_{cd}
\eeq
for $A_\mu^{a}$.
Integrating out the auxiliary field, we reproduce the original
action (\ref{originalaction}).

Since the $A_\mu ^{~a}$ is quadratic in 
(\ref{auxiliaryaction}), we can integrate it out first.
The propagator for $A_\mu^{~a}$ is given by
\beq
\langle A_\mu^{~a} A_\nu^{~b} \rangle
= g \delta_{\mu\nu} (K^{-1})_{ab} .
\eeq
For example, $\langle \tr (A^2) \rangle$ can be 
expressed by the following integral.
\beq
\langle \tr(A^2) \rangle
= 
\frac{ \int \dd h_{ab} g D (K^{-1})_{aa} \ee^{- S_{\mbox{\tiny
eff}}} }
{ \int \dd h_{ab} \ee^{- S_{\mbox{\tiny eff}}} } .
\eeq
$S_{\mbox{\scriptsize eff}}$ is 
the effective action for $h$ defined by
\beq
S_{\mbox{\scriptsize eff}}
=
 \frac{D}{2} \Tr \ln K + 
\frac{1}{4 g^2} \lambda ^{abcd} h_{ab} h_{cd} ,
\eeq
where the $\Tr$ represents a trace over 
the adjoint indices $a$.

By rescaling the $h$ as
$\tilde{h}_{ab} = \frac{1}{g \sqrt{D}} h_{ab}$,
which is now dimensionless, we can rewrite the effective action
as
\beq
S_{\mbox{\scriptsize eff}}
=
 \frac{D}{2} \left \{ \Tr \ln K + 
\frac{1}{2} \lambda ^{abcd} \tilde{h}_{ab} \tilde{h}_{cd} \right \} .
\eeq
Therefore, in the large $D$ limit, the integration over $\tilde{h}$ 
is dominated by the saddle point and one can perform
a systematic $1/D$ expansion.
The saddle point equation can be given as
\beq
(K^{-1})_{cd} \frac{\delta K_{dc}}{\delta \tilde{h}_{ab}}
+ \lambda ^{abcd} \tilde{h}_{cd} = 0 .
\label{speq}
\eeq
The relevant saddle point $\tilde{h}_{ab}^{(0)}$,
if we assume that it preserves the SU($N$) symmetry,
should be written as
$\tilde{h}_{ab}^{(0)} = v \delta_{ab}$.
From the saddle point equation (\ref{speq}),
we obtain $v=\frac{1}{\sqrt{2N}}$.
We expand the $\tilde{h}_{ab}$ around the saddle point
$\tilde{h}_{ab}^{(0)}$ as
\beq
\tilde{h}_{ab} = \frac{1}{\sqrt{2N}} \delta_{ab} 
+ 2 \sqrt{\frac{N}{D}}\varphi _{ab},
\eeq
where $\varphi _{ab}$ is real symmetric.
We have put the factor $2 \sqrt{\frac{N}{D}}$
in front of $\varphi _{ab}$ for later convenience.
Expanding $K$ in terms of $\varphi$, we obtain
\beq
K_{ab}= \sqrt{2ND} (\delta_{ab} - \epsilon X_{ab}),
\eeq
where $\epsilon = \sqrt{\frac{2}{D}}$ and
\beq
X_{ab} = \lambda^{abcd} \varphi_{cd}.
\eeq
We can expand $(K^{-1})_{ab}$ and $\Tr \ln K$
in terms of $\varphi$ in the following way.
\beqa
(K^{-1})_{ab} &=& 
\frac{1}{\sqrt{2ND}}
(\delta_{ab}  + \epsilon X_{ab} + \epsilon ^2 
X_{ac}X_{cd} + O(\varphi^3 ))  , \\
\Tr \ln K &=&
\mbox{const.} - \epsilon X_{aa} - 
\frac{1}{2} \epsilon ^2  X_{ab}X_{ba} + O(\varphi^3 ) .
\eeqa
The effective action $S_{\mbox{\scriptsize eff}}$ can be expanded
in terms of $\varphi$ as
\beq
S_{\mbox{\scriptsize eff}}
= \frac{1}{2}
T^{abcd}\varphi_{ab}\varphi_{cd} 
- \sum_{n=3}^{\infty} \frac{\epsilon ^{n-2}}{n} \Tr (X^n) 
+ \mbox{const.},
\eeq
where we have defined
\beq
T^{abcd} =  - \lambda^{abef} \lambda^{cdef} 
+ 2 N \lambda^{abcd} .
\eeq

The propagator of $\varphi_{ab}$ can be given by
\beq
\langle \varphi _{ab} \varphi _{cd} \rangle
=  S^{abcd},
\eeq
where $S^{abcd}$ is defined by
\beq
T^{abpq}S^{pqcd} = S^{abpq}T^{pqcd} = I^{abcd}.
\eeq
In order to obtain $S^{abcd}$ explicitly,
we introduce the following quantities.
\beqa
F^{abcd} &=& \frac{1}{4} (f^{abcd}+f^{bacd}+f^{abdc}+f^{badc}) ,
\label{defF}
\\
G^{abcd} &=& \frac{1}{2} (f^{acbd}+f^{adbc}) ,
\label{defG}
\\
H^{abcd} &=& \delta_{ab} \delta_{cd} ,
\label{defH}
\\
I^{abcd} &=& \frac{1}{2}(\delta_{ac}\delta_{bd}
+\delta_{ad}\delta_{bc}),
\label{defI}
\eeqa
where
\beq
f ^{abcd} = \tr ( t^a t^b t^c t^d ) .
\eeq
$\lambda^{abcd}$ and $T^{abcd}$ can be written as follows.
\beqa
\lambda ^{abcd} &=& 2 (G^{abcd}- F^{abcd}) \\
T^{abcd} &=& - 6N F^{abcd} + 4N G^{abcd} - 2 H^{abcd} - 4 I^{abcd}.
\eeqa
$S^{abcd}$ can be obtained as
\beq
S^{abcd}= \frac{1}{4(N^2 -1 )}
\left\{ 
- \frac{2}{3N} F^{abcd}  + N G^{abcd}
+ \frac{1}{6 N^2} H^{abcd} + I^{abcd}
\right\}.
\eeq

\begin{figure}[htbp]
  \begin{center}
    \includegraphics[height=3cm]{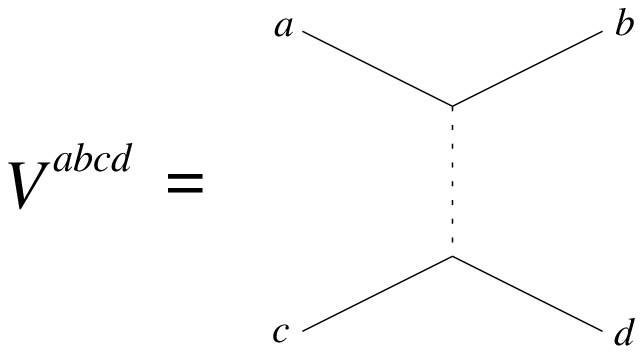}
    \caption{}
    \label{fig:Vabcd}
  \end{center}
\end{figure}

Since we work with $X_{ab}$ instead of $\varphi_{ab}$ in what follows,
we introduce the propagator of $X_{ab}$ given as
\beqa
V^{abcd} &\defeq& \langle X_{ab} X_{cd} \rangle \\
&=&
\lambda^{abef} S^{efgh} \lambda^{ghcd} \\
&=& \frac{1}{N^2-1}
\left\{
- \frac{2}{3} N F^{abcd} + N G^{abcd}
+ \frac{1}{6} H^{abcd} + I^{abcd} \right\}.
\label{Vdef}
\eeqa

\begin{figure}[htbp]
  \begin{center}
    \includegraphics[height=6.5cm]{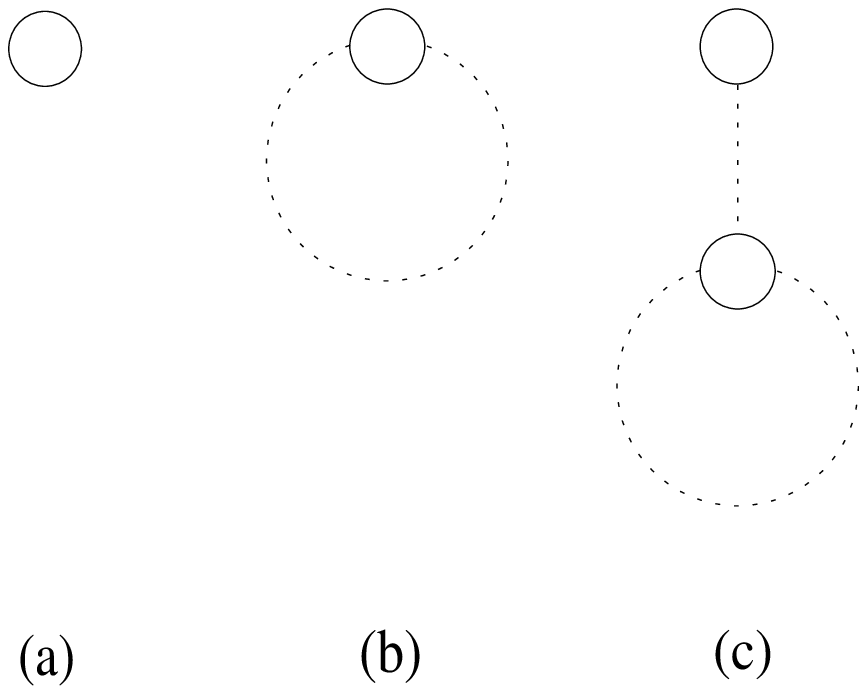}
    \caption{}
    \label{fig:tra2}
  \end{center}
\end{figure}

Let us represent $V^{abcd}$ by the four-point vertex,
as is shown in Fig. \ref{fig:Vabcd}.
Then we can calculate
$\langle \tr (A^2) \rangle$ by evaluating
the diagrams depicted in Fig. \ref{fig:tra2}.
The result is given as follows.
\beq
\langle \tr (A^2) \rangle
= g D \frac{1}{\sqrt{2ND}} \left\{
(N^2 -1 ) + \epsilon ^2 \left( 
V^{abab}+ V^{aabc} V^{bddc} \right)
+ O(\epsilon ^4) \right\},
\label{diagramA2}
\eeq
where each term corresponds to the diagrams (a), (b) and (c)
in Fig. \ref{fig:tra2}, respectively.
By noting that $V^{abcc}= -\frac{1}{2}\delta_{ab}$ and
$V^{abab}= \frac{1}{6}(7 N^2 -1)$,
we obtain the result for the quantity (1) defined in 
Section \ref{monte} as
\beqa
(1) &\defeq&
\left\langle \frac{1}{N} \tr (A^2) \right\rangle
\times \left( \sqrt{g}N^{1/4} \right)^{-2} \n
&=& \sqrt{\frac{D}{2}} \left\{ \left(1 - \frac{1}{N^2}\right) + 
\frac{1}{D}\left(\frac{7}{6} - \frac{1}{6 N^2}\right) 
+ O \left(\frac{1}{D^2}\right)\right\}.
\label{op1sub}
\eeqa

Similarly, we can calculate 
$\langle \tr ((A_\mu A_\nu)^2) \rangle$ as follows.
\beqa
&~&\langle \tr ((A_\mu A_\nu)^2) \rangle \n
&=& G^{abcd} \langle A_\mu^{~a} A_\mu^{~b} A_\nu^{~c} A_\nu^{~d} 
\rangle \n
&=& g^2 G^{abcd} \{ D^2 \langle (K^{-1})_{ab} (K^{-1})_{cd}\rangle +
2 D \langle(K^{-1})_{ac} (K^{-1})_{bd}\rangle  \} \n
&=& g^2 \frac{1}{2ND} 
[ D^2 \{ G^{aabb} + \epsilon ^2
 (G^{abcd}V^{abcd} + 2 G^{aabc}V^{bddc}
+ 2 G^{abcc} V^{abde} V^{dffe} )\} \n
&~& ~~~~~~~~~~+ 2 D G^{abab}  + O(1)].
\eeqa
$\langle \tr ((A^2)^2) \rangle$ can be calculated by replacing 
$G$ by $F$ in the above expressions.
We obtain the following final results for the quantities (2) and (3).
\beqa
(2) &\defeq &  \left\langle 
\frac{1}{N} 
 \tr ((A_\mu A_\nu)^2) \right\rangle 
\times \left( \sqrt{g}N^{1/4} \right)^{-4} \n
&=& D \left\{ - \frac{1}{2 N^2}
\left(1 - \frac{1}{N^2}\right) + 
\frac{1}{D}\left(\frac{3}{2} - \frac{5}{2 N^2}
+\frac{1}{N^4} \right) + O \left(\frac{1}{D^2}\right)\right\} ,
\label{op2sub}
\\
(3) &  \defeq  & \left\langle 
\frac{1}{N} 
 \tr ((A^2)^2) \right\rangle 
\times \left( \sqrt{g}N^{1/4} \right)^{-4} \n
&=& 
D \left\{ \frac{1}{2}
\left(1 - \frac{1}{N^2}\right)^2 + 
\frac{1}{D}\left(\frac{3}{2} - \frac{5}{2 N^2}
+\frac{1}{N^4} \right) + O \left(\frac{1}{D^2}\right)\right\}  .
\label{op3sub}
\eeqa
Note that the above results are consistent with
the exact result (\ref{trf2}) for
$\langle \frac{1}{N} \tr (F^2) \rangle $.

\subsection{Large $N$ behavior and factorization}
\label{largeNfactorization}

We estimate the order in $N$ of correlation functions
to all orders of the $1/D$ expansion
and further prove the large $N$ factorization property.

Let us first consider $\langle \frac{1}{N}\tr (A^2) \rangle$
as an example.
The diagrams we encounter up to the next-leading order
in the $1/D$ expansion are shown in Fig. \ref{fig:tra2}.
We estimate the order in $N$ of each diagram we encounter
at each order of the $1/D$ expansion.
We start by replacing the $V^{abcd}$ by the r.h.s. of (\ref{Vdef}).
Each $V$-vertex should be replaced by either of
the $F$-, $G$-,$H$- and $I$-vertices with the factor
$\frac{1}{N}$ for the first two, 
and with the factor $\frac{1}{N^2}$ for the last two
due to the coefficients in (\ref{Vdef}).
We denote the number of $F$- and $G$-vertices by $V_1$ 
and the number of $H$- and $I$-vertices by $V_2$
in the diagram after this replacement.
The diagram has an overall factor
$\left(\frac{1}{N} \right)^{V_1}
\left(\frac{1}{N^2} \right)^{V_2}$.

\begin{figure}[htbp]
  \begin{center}
    \includegraphics[height=3cm]{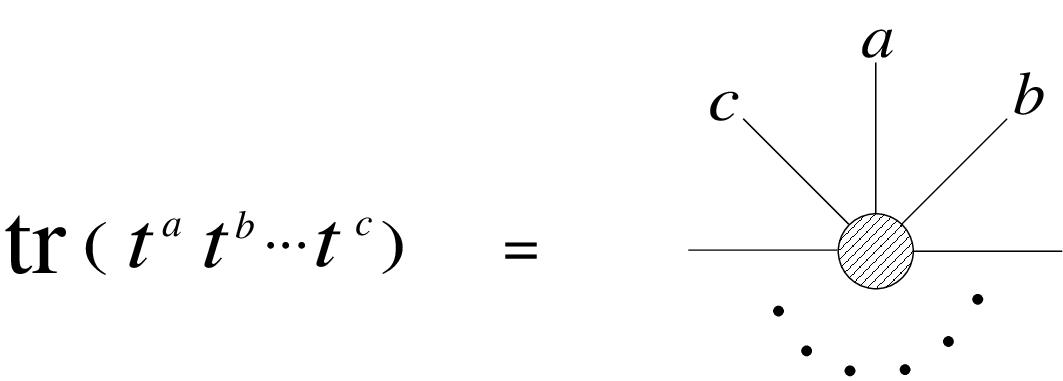}
    \caption{}
    \label{fig:Tabcd}
  \end{center}
\end{figure}

\begin{figure}[htbp]
  \begin{center}
    \includegraphics[height=5cm]{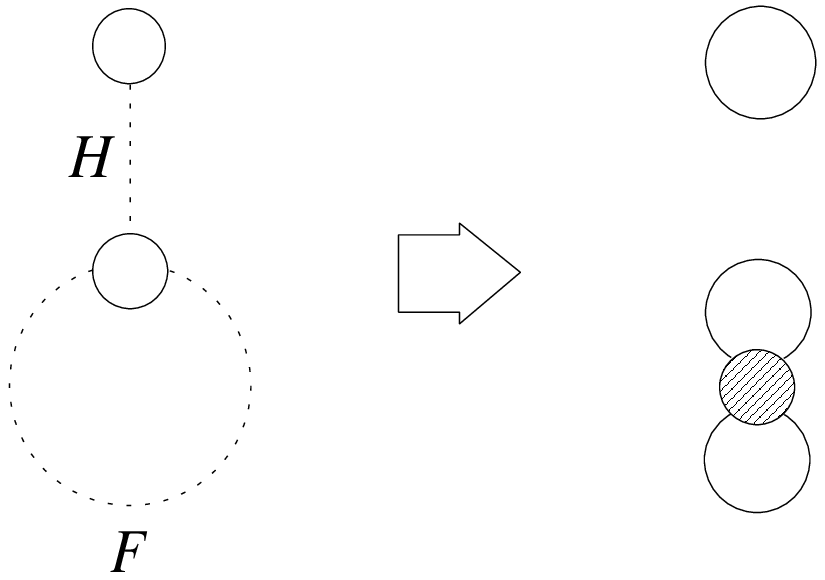}
    \caption{}
    \label{fig:typ}
  \end{center}
\end{figure}

The next task is to replace the 
$F$-,$G$-,$H$- and $I$-vertices by their definitions
(\ref{defF})$\sim$(\ref{defI}).
Let us introduce the diagrammatic representation for
$\tr (t^a t^b \cdots t^c )$ as in Fig. \ref{fig:Tabcd}.
Fig. \ref{fig:typ} shows an example of what we get
from the diagram (c) of Fig. \ref{fig:tra2}.
The symbols $F$ and $H$ in the l.h.s. of Fig. \ref{fig:typ}
represent that the $V$-vertices in the diagram (c) of Fig. \ref{fig:tra2}
have been replaced by $F$- and $H$- vertices respectively.
In general, we obtain diagrams composed of 
$V_1$ blobs connected by $L$ lines 
with additional $\ell$ loops which are not connected to any of the
blobs.
Reflecting the fact that each blob
has four legs, we have 
\beq
4 V_1 = 2 L .
\label{LtoV1}
\eeq
Let us denote by $C$ the number of connected parts in the diagram
excluding the additional $\ell$ loops.
For the diagram shown in the r.h.s. of Fig. \ref{fig:typ},
we have $C=1$ and $\ell = 1$.
Since the number of connected parts has a chance to increase
from its initial value `1' by the use of $H$- and $I$- vertices,
we have
\beq
C+\ell = 1+ j ,
\label{ell}
\eeq
where $j=0,1,\cdots,V_2$.
The $\ell$ loops give a factor of $(N^2)^\ell$.

\begin{figure}[htbp]
  \begin{center}
    \includegraphics[height=3cm]{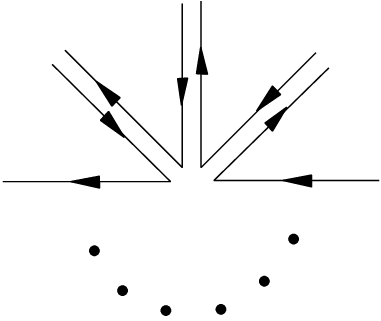}
    \caption{}
    \label{fig:doubleT}
  \end{center}
\end{figure}

\begin{figure}[htbp]
  \begin{center}
    \includegraphics[height=1cm]{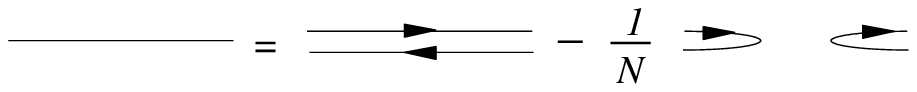}
    \caption{}
    \label{fig:contract}
  \end{center}
\end{figure}

\begin{figure}[htbp]
  \begin{center}
    \includegraphics[height=6cm]{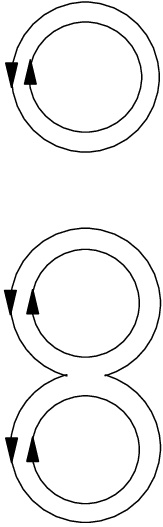}
    \caption{}
    \label{fig:typ_double}
  \end{center}
\end{figure}

In order to read off the order in $N$ of the diagram, it is convenient
to switch at this stage to the double-line notation.  The $\tr (t^a
t^b \cdots t^c )$ should be represented as in Fig. \ref{fig:doubleT}.
Due to the relation (\ref{Trformula}), we have the diagrammatic rule
depicted in Fig. \ref{fig:contract}
for the contraction of the adjoint indices.  
In order to see the leading large $N$ behavior,
it suffices to consider only the first term.
%One can easily see that
%the second term can be neglected as far as leading large $N$ 
%contributions are concerned.
Fig. \ref{fig:typ_double} shows the diagram in the double-line notation
which gives the leading large $N$ contribution to the diagram on the
r.h.s. in Fig. \ref{fig:typ}.
If we denote by $F$ the number of index loops
excluding those coming from the $\ell$ additional loops,
we have the following relation
\beq
F+V_1 - L = 2C -2h,
\label{Euler}
\eeq
where $h$ is the number of handles necessary to write
the diagram without crossings of the double lines.

The order of magnitude of the contribution to 
$\langle \frac{1}{N} \tr (A^2) \rangle$
from the diagram is given by
\beqa
&~& \frac{1}{N}\frac{g}{\sqrt{N}} \left( \frac{1}{N} \right) ^{V_1} 
    \left( \frac{1}{N^2} \right) ^{V_2} 
    (N^2)^\ell N^F \n
&=& g\sqrt{N} \cdot (N^2)^{(j- V_2) - h},
\label{trA2exp}
\eeqa
where we have used 
(\ref{LtoV1}),(\ref{ell}) and (\ref{Euler}).
Therefore the leading term is $O(g \sqrt{N})$, which is obtained for
$j= V_2$ and $h=0$.

So far, we have been concentrating on the leading large $N$ behavior.
%of the quantity $\langle \frac{1}{N} \tr (A^2) \rangle$.
Let us next consider sub-leading terms.
Note first that the sub-leading terms from (\ref{trA2exp}) are
given in terms of a $1/N^2$ expansion.
The effect of the second term in
Fig. \ref{fig:contract}
is a factor either of $O(1)$ or of $O(\frac{1}{N^2})$
for each replacement of the first term by the second term.
Also the prefactor $\frac{1}{N^2-1}$ of
(\ref{Vdef}) can be expanded in terms of $1/N^2$, 
which may give extra $O(\frac{1}{N^2})$ factors, as well.
Thus we have shown that
the subleading terms appear as a $1/N^2$ expansion.

Generalization of the above argument to the quantity
written as
\beq
\left \langle \frac{1}{N}\tr (\underbrace {A\cdots A}_{p_1} )
\frac{1}{N}\tr (\underbrace {A\cdots A}_{p_2} )
\cdots
\frac{1}{N}\tr (\underbrace {A\cdots A}_{p_Q}) \right\rangle
\label{generaltra}
\eeq
is straightforward.
(\ref{LtoV1}), (\ref{ell}) and (\ref{Euler})
should be replaced by\footnote{(\ref{LtoV1_general}), 
(\ref{ell_general}) and (\ref{Euler_general}) 
do not reduce to
(\ref{LtoV1}), (\ref{ell}) and (\ref{Euler}),
when we take $Q=1$, $p_1=2$.
This is simply because we have omitted the blob coming from 
the trace of $\langle \tr (A^2) \rangle$ using
the fact that $\tr (t^a t^b) = \delta_{ab}$.
The general argument below holds for the particular case as well.}
\beqa
4 V_1 + P &=& 2 L ,
\label{LtoV1_general}
\\
C + \ell &=& q + j, 
\label{ell_general}
\\
F + (V_1 + Q) - L &=& 2 C - 2 h,
\label{Euler_general}
\eeqa
where $P=\sum_{i=1}^{Q}  p_i $ and $q=1,2,\cdots, Q$.
$q$ is the number of connected parts before replacing the
$V$-vertices by either of $F$-,$G$-,$H$- and $I$-vertices.
Then the order of magnitude of
the quantity (\ref{generaltra}) is given by
\beqa
&~& \left(\frac{1}{N}\right)^Q
  \left(\frac{g}{\sqrt{N}}\right)^{P/2} 
\left( \frac{1}{N} \right) ^{V_1} 
    \left( \frac{1}{N^2} \right) ^{V_2} 
    (N^2)^\ell N^F \n
&=& (\sqrt{g} N^{1/4})^P N^{ 2\{(q-Q)+(j-V_2)-h \}} ,
\label{generalres}
\eeqa
where we have used (\ref{LtoV1_general}),
(\ref{ell_general}) and (\ref{Euler_general}).
Therefore, we have the maximum order 
$(\sqrt{g} N^{1/4})^P$
when $q=Q$, $j=V_2$ and $h=0$
and the sub-leading terms appear as a $1/N^2$ expansion.

Thus we have shown, to all orders of the $1/D$ expansion,
that if we define
\beq
{\cal O}= 
\frac{1}{N} \tr (\underbrace {A\cdots A}_{p} )
\times (\sqrt{g}N^{1/4})^{-p},
\eeq
correlation functions of ${\cal O}$'s 
such as $\langle {\cal O}_1 {\cal O}_2 \cdots {\cal O}_n \rangle$
have finite large $N$ limits.
Note that this large $N$ behavior is the same as
the one obtained by the perturbative argument
in Section \ref{perturb} with the input
of $R \sim \sqrt{g} N ^{1/4}$.
The fact that the large $N$ behavior of the correlation functions 
does not depend on the order of the $1/D$ expansion
is consistent with the fact that
the perturbative argument is independent of $D$
except for $D=3$.
We have also shown that the subleading large $N$ 
behavior of the correlation functions can be given by a
$1/N^2$ expansion,
which is clearly seen by the Monte Carlo simulation
in Section \ref{monte}.
Moreover, the fact that the leading large $N$ contribution 
to (\ref{generaltra}) comes
solely from the diagrams with $q=Q$
as is seen from (\ref{generalres})
means that we have the factorization property : 
\beq
\langle {\cal O}_1 {\cal O}_2 \cdots {\cal O}_n \rangle
= \langle {\cal O}_1 \rangle \langle {\cal O}_2 \rangle
\cdots \langle  {\cal O}_n \rangle + O\left(\frac{1}{N^2}\right)
\label{factorprop}
\eeq
in the large $N$ limit.

The factorization can be generalized
to the case when ${\cal O}$'s are Wilson loop operators such as
\beq
{\cal O}=
\frac{1}{N}
{\cal P} \exp (i\int \dd \sigma k_\mu (\sigma) A_\mu) .
\eeq
We here consider $k_\mu (\sigma)$ to be written as
\beq
k_\mu (\sigma) = \frac{1}{\sqrt{g}N^{1/4}} f_\mu (\sigma),
\eeq
where $f_\mu (\sigma)$ is independent of $g$ and $N$.
Then the above quantity can be expanded in terms of $f$ as
\beqa
{\cal O}&=&
\frac{1}{N}
{\cal P} \exp (i\int \dd \sigma k_\mu (\sigma) A_\mu) \n
&=&\sum_{n=0}^{\infty} i^n
\int_{0}^{1} \dd \sigma _1
\int_{\sigma_1}^{1} \dd \sigma _2
\cdots
\int_{\sigma_{n-1}}^{1}\dd \sigma _n
f_{\mu_1} (\sigma_1)
f_{\mu_2} (\sigma_2)
\cdots
f_{\mu_n} (\sigma_n)  \n
&~& ~~~\frac{1}{N}
\tr (A_{\mu_n}\cdots A_{\mu_2}A_{\mu_1})
\times (\sqrt{g} N^{1/4})^{-n}.
\eeqa
Therefore the factorization property is satisfied by 
the Wilson operators considered above as well.
This is in contrast to the situation seen in the double scaling 
limit of the 2D Eguchi-Kawai model,
studied as a toy model of the IIB matrix model \cite{NN}.

\subsection{Comparison of the $1/D$ expansion with numerical data}
\label{comparison}

\begin{figure}[htbp]
  \begin{center}
    \includegraphics[height=10cm]{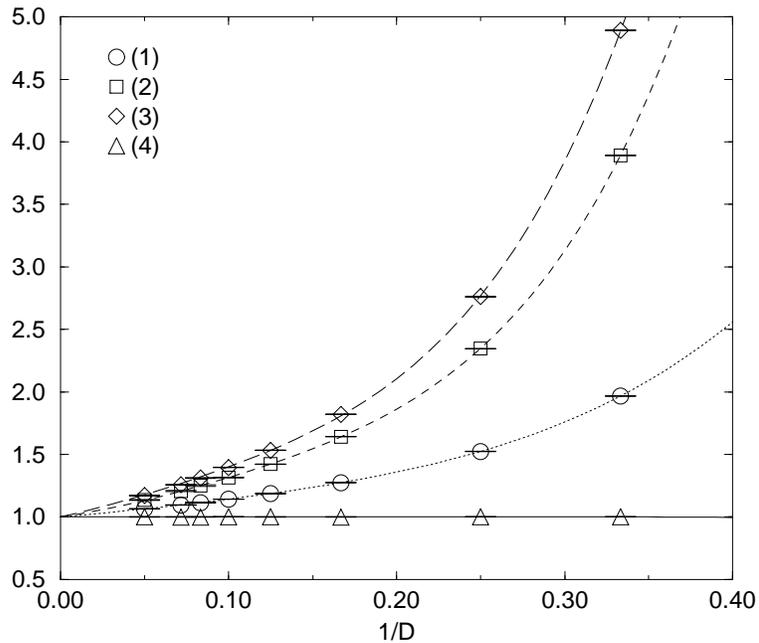}
    \caption{
      The quantities (1)(circles),(2)(squares),(3)(diamonds) and
      (4)(triangles) extrapolated to $N=\infty$ and normalized by the
      leading asymptotic behavior in the large $D$ limit are plotted
      against $1/D$ for the bosonic model.  The solid line represents
      the exact result (\ref{trf2}) for (4) and the other lines
      represent the fits to $1 + b_1/D + b_2/D^2 + b_3/D^3 + b_4/D^4$
      with the input of the $1/D$ expansion up to the next-leading
      terms.}
    \label{fig:rN}
  \end{center}
\end{figure}

In order to confirm that the $1/D$ expansion
developed in the previous sections is valid for studying
the large $N$ limit of the model for finite $D$,
we perform Monte Carlo simulations for various $D$
and examine the $D$ dependence of the quantities (1)$\sim$(4)
defined in Section \ref{monte}.
Extrapolation to $N=\infty$ has been done by
the fit to $a_0 + a_1/N^2$ 
with $N=16$, 32, 64, 128, 256
for $D=3$, 4,
with $N=16$, 32, 64 for $D=6$, 8,
and with $N=16$, $32$ for $D=10,12,14,20$.
In order to compare the data with the results of the $1/D$ expansion
(\ref{op1sub}), (\ref{op2sub}) and (\ref{op3sub}),
where we put $N =\infty$,
we normalize the Monte Carlo data by the leading 
asymptotic behavior in the large $D$ limit.
For (2), since the leading term in $1/D$ vanishes in the 
$N \rightarrow \infty $ limit,
we normalize the data by the next-leading term.
In Fig. \ref{fig:rN}, we plot the results against $1/D$.
We fit the data to $1 + b_1/D + b_2/D^2 + b_3/D^3 + b_4/D^4$,
except for (4), which is known exactly.
For (1) and (3), we use the values of $b_1$ 
predicted by the next-leading term.
We find that the data are nicely fitted.

This result suggests that 
we do not have a phase transition at some intermediate value
of $D$ and that the $1/D$ expansion is valid 
for finite $D$ down to $D=3$.

\vspace*{1cm}

\section{No SSB of Lorentz invariance}
\setcounter{equation}{0}
\label{lorentzSSB}

In this section, we address the issue of the SSB of the Lorentz
invariance in large $N$ reduced models.
This is of paramount importance in the IIB matrix model,
since if the space time is to be four-dimensional at all,
the 10D Lorentz invariance of the model must be spontaneously broken.

Let us define an order parameter
for the SSB of the Lorentz invariance.
We first note that the extent of the eigenvalue distribution
in the direction $n_{\mu}$, where $n_{\mu}$ is a $D$-dimensional
unit vector, can be given by the square root of 
\beqa
I(n) &=& \frac{1}{N} \tr ~ (n\cdot A)^2 \\
     &=& I_{\mu\nu} n_\mu n_\nu ,
\eeqa
where $I_{\mu\nu}$ is a $D \times D$ real symmetric matrix defined
by
\beq
I_{\mu\nu} = \frac{1}{N} \tr (A_\mu A_\nu).
\eeq
The eigenvalues of $I_{\mu\nu}$ are real positive.
The SSB of the Lorentz invariance can be probed
by the variation of the eigenvalues, which 
is given by
\beqa
J &=& \frac{1}{D} I_{\mu\nu} I_{\mu\nu} -
        \left( \frac{1}{D} I_{\mu\mu} \right)^2 \\
 &=& \frac{1}{D} 
\left\{ \frac{1}{N} \tr(A_\mu A_\nu)  \right\}^2
-  \frac{1}{D^2} \left\{ \frac{1}{N} \tr(A ^2 )  \right\}^2.
\eeqa
If $\langle J \rangle/ R^4 $ 
is nonzero in the large $N$ limit,
where $R\sim \sqrt{g}N^{1/4}$
is the (averaged) extent of space time,
the Lorentz invariance is spontaneously broken.
Therefore, $\langle J \rangle/ (\sqrt{g}N^{1/4})^4 $ 
can be considered as an order parameter 
for the SSB of the Lorentz invariance.

\begin{figure}[htbp]
  \begin{center}
    \includegraphics[height=10cm]{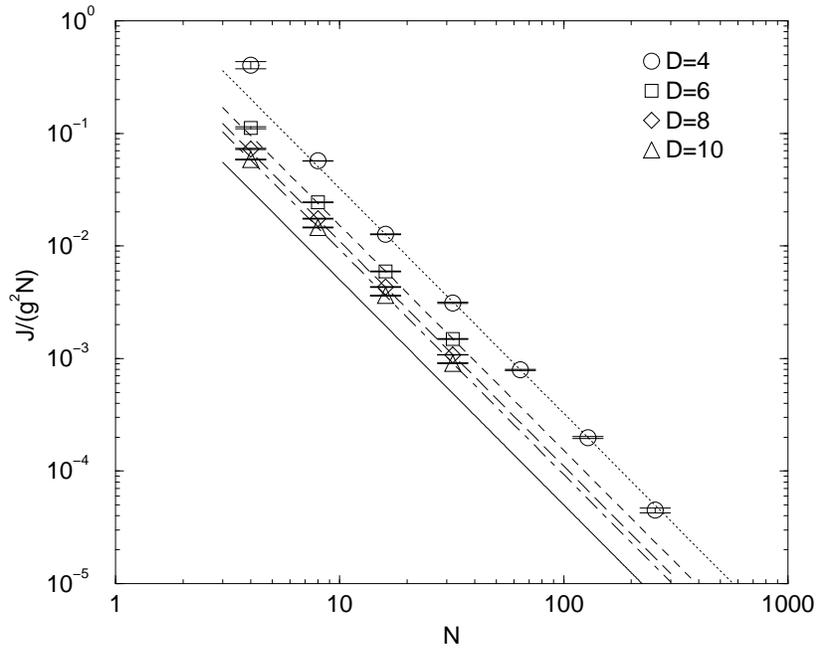}
    \caption{The order parameter for the SSB of the Lorentz invariance
      $\langle J \rangle/ (\sqrt{g}N^{1/4})^4 $ is plotted against $N$
      for $D=4,6,8$ and $10$.  The solid line represents the leading
      large $N$ behavior for $D=\infty$ and the other lines represent
      the fits of the data to the $1/N^2$ behavior predicted by 
      (\ref{factorprop}).}
    \label{fig:5D}
  \end{center}
\end{figure}

In Fig. \ref{fig:5D} we plot $\frac{\langle J \rangle}
{(\sqrt{g}N^{1/4})^4}$ 
against $N$ for $D=4,6,8$ and $10$.
We see that the order parameter vanishes in the large $N$ limit,
which means that the Lorentz invariance is not spontaneously broken.

%We point out that 
This result can be also confirmed by the $1/D$ expansion.
By using the $1/D$ expansion, we can prove a relation
\beq
\langle  \tr(A_\mu A_\nu)  \rangle = 
\frac{1}{D} \delta_{\mu\nu}
\langle  \tr(A^2)  \rangle ,
\eeq
which already suggests the absence of an SSB of 
the Lorentz invariance.
%as a direct consequence of 
%the large $N$ factorization.
In fact, this relation combined with the relation
\beqa
&~& \left \langle  \frac{1}{N}
\tr(A_\mu A_\nu) \frac{1}{N}\tr(A_\lambda A_\rho) \right\rangle
\times \left( \sqrt{g}N^{1/4} \right)^{-4}  \n
&=&  \left\langle  \frac{1}{N}\tr(A_\mu A_\nu)  \right \rangle 
\left\langle \frac{1}{N}\tr(A_\lambda A_\rho) \right\rangle 
\times \left( \sqrt{g}N^{1/4} \right)^{-4} 
+ O \left(\frac{1}{N^2}\right),
\label{eq:factorization}
\eeqa
due to the large $N$ factorization property (\ref{factorprop}),
which can be also shown by the $1/D$ expansion,
leads to $\frac{\langle J \rangle}{(\sqrt{g}N^{1/4})^4}
 = O(\frac{1}{N^2})$ in the large $N$ limit.
%By using
%\beq
%\langle  \tr(A_\mu A_\nu)  \rangle = 
%\frac{1}{D} \delta_{\mu\nu}
%\langle  \tr(A^2)  \rangle ,
%\eeq
%which can be shown by the $1/D$ expansion,
%Therefore we obtain $\frac{\langle J \rangle}{(\sqrt{g}N^{1/4})^4}
% = O(\frac{1}{N^2})$ in the large $N$ limit.
Indeed the large $N$ behavior of the order parameter obtained
by the Monte Carlo simulation
can be nicely fitted to $1/N^2$.

We have also calculated the order parameter by the $1/D$ expansion
explicitly.
The result is given as
\beq
\frac{\langle J \rangle }{(\sqrt{g}N^{1/4})^4}=
\frac{1}{2 N^2} \left\{ 1+\frac{17}{3D}
+ O \left(\frac{1}{D^2} \right) \right\}+
O\left(\frac{1}{N^4}\right).
\eeq
The leading large $N$ behavior for $D=\infty$
is plotted in Fig. \ref{fig:5D} for comparison.

\vspace*{1cm}

\section{Summary and Discussion}
\setcounter{equation}{0}
\label{summary}

In this paper, we studied the reduced model of bosonic Yang-Mills
theory with special attention to dynamical aspects related to the
eigenvalues of the $N \times N$ matrices, which are regarded as the
space-time coordinates in the IIB matrix model.  We found that the
bosonic model allows a systematic $1/D$ expansion, which indeed
revealed many interesting dynamical properties of the model.  One
should note again that the parameter $g$ of the model, which
corresponds to the coupling constant before being reduced, is merely a
scale parameter which can be scaled out if one wishes by an
appropriate redefinition of the dynamical variables $A_\mu$.  Thus,
unlike in ordinary gauge theories, we have neither a weak
coupling expansion nor a strong coupling expansion.  $1/D$ is the only
parameter which allows a systematic expansion of the model.  We
confirmed that the $1/D$ expansion is indeed valid for all $D \ge 3$.

To all orders of the expansion, we estimated the large $N$ behavior of
correlation functions, which agrees with the one determined by the
Monte Carlo simulation.  Above all, the extent of space time, which is
defined by $R=\sqrt{\langle \frac{1}{N} \tr(A^2) \rangle}$,
was found to behave as $R \sim \sqrt{g} N^{1/4}$ in the large $N$ limit.
For $D\ge 4$,
the above result turned out to saturate
the upper bound given
by perturbative arguments, which fact makes the perturbative
estimation of the large $N$ behavior of correlation functions valid.
We also showed that the large $N$ behavior of the
extent of space time is consistent with the SSB of the U(1)$^D$
symmetry in the weak coupling region of the Eguchi-Kawai model.

The fact that we could obtain an upper bound on $R$ through
a perturbative evaluation of the effective action
has an important implication.  It means
that if we look at the effective action for the eigenvalues
$\lambda_{\mu i}$ obtained perturbatively, 
it has nonnegligible contributions when
a pair of $\lambda_i$'s which are assigned to adjacent index loops on a
Feynman diagram ({\it e.g.}, Fig. \ref{fig:2loop}) are separated
widely in the $D$-dimensional space time.  This gives an obstacle to
identifying the Feynman diagrams with smooth worldsheets of strings,
which means that the bosonic model cannot be interpreted as a string
theory.  We expect that the situation changes drastically in the
supersymmetric case.
% \cite{NT}.

We defined a quantity representing the uncertainty of the space-time
coordinates and showed that it is of the same order as the extent of
space time, which means that a classical space-time picture is
maximally broken.
To all orders of the $1/D$ expansion, we found that correlation
functions are given by an expansion in terms of $1/N^2$.  This makes,
{\it e.g.}, $N=16$ 
sufficiently large to probe the large $N$ behavior of the model
even for large $D$.  We pointed out that this is related to the 
maximal breakdown of
the classical space-time picture.

%We proved the large $N$ factorization property of correlation
%functions to all orders of the $1/D$ expansion.  
The absence of an SSB of
the Lorentz invariance was shown both by the Monte Carlo simulation
and by the $1/D$ expansion.
%as a direct consequence of the large
%$N$ factorization.  
%Note that the $1/D$ expansion cannot be formulated
%in the supersymmetric case, since $D$ should be kept fixed to preserve
%supersymmetry.  This makes nontrivial, above all, 
%whether the large $N$ factorization holds as well
%in the supersymmetric case,
%which opens up a possibility of the SSB of the Lorentz
%invariance.

All the above issues we addressed for the bosonic model should be
addressed for the IIB matrix model as well.  The results must reveal
all the dynamical properties of the model related to the space-time
structure and the worldsheet picture.  We expect that the
supersymmetry plays an essential role.  We hope our findings for the
bosonic model provide a helpful comparison when we investigate the
supersymmetric case including the IIB matrix model.

\vspace*{1cm}

\section*{Acknowledgements}
We would like to thank H. Aoki and H. Kawai for stimulating
discussions and helpful communications.  We are also grateful to N.
Ishibashi, S. Iso, H. Itoyama, Y. Kitazawa, T. Nakajima, T. Suyama, T.
Tada and T. Yukawa for discussions.  This work is supported by the
Supercomputer Project (No.98-38) of High Energy Accelerator Research
Organization (KEK).  The work of J.N. and A.T. is supported in part by
Grant-in-Aid for Scientific Research (Nos. 10740113 and 10740121) from
the Ministry of Education, Science and Culture.

\newpage

\section*{Appendix: The algorithm for the Monte Carlo simulation}
\setcounter{equation}{0}
\renewcommand{\theequation}{A.\arabic{equation}}
\hspace*{\parindent}
In this appendix, we explain the algorithm we use
for the Monte Carlo simulation of the bosonic model
with the action
\beq
S=-\frac{1}{2} \beta  \sum_{\mu < \nu} 
\tr([A_{\mu},A_{\nu}]^2),
\label{bosonicaction}
\eeq
where $\beta = \frac{1}{g^2}$.
Since $\mu \neq \nu$, the action is quadratic 
with respect to each component,
which means that we can 
update each component by generating gaussian variables
using the heat bath algorithm.

We first note that the trace can be written as
\beq
\tr([A_{\mu},A_{\nu}]^2)
= 2 (A_\mu)_{ij} (A_\nu)_{jk}  (A_\mu)_{kl}  (A_\nu)_{li} 
  - 2 (A_\mu)_{ij} (A_\nu ^{~2})_{jk}  (A_\mu)_{ki}.
\eeq 
Suppose we want to update $A_\mu$.
In the first term, 
all the $N^2$ components are coupled,
while in the second term, only the components which include
at least one common index are coupled.

In order to simplify the algorithm, 
we employ the following trick.
We first rewrite the action as
\beq
S=-\frac{1}{2} \beta 
\sum_{\mu < \nu} \tr(\{A_{\mu},A_{\nu}\} ^2 )
+ 2 \beta \sum_{\mu < \nu} \tr(A_{\mu} ^{~2} A_{\nu} ^{~2}  ).
\label{bosonicaction2}
\eeq
Here we introduce the auxiliary field
$Q_{\mu\nu}$ ($1 \le \mu < \nu \le D $),
which is $~_D \mbox{C} _2$ $N \times N$ hermitian matrices,
with the following action
\beq
S' = \frac{1}{2} \beta
\sum_{\mu < \nu}
\tr (Q_{\mu\nu} ^{~2}   - 2 Q_{\mu\nu} G_{\mu\nu}  )
+ 2 \beta \sum_{\mu < \nu} \tr (A_\mu ^{~2} A_\nu ^{~2}),
\label{eq:vectorize}
\eeq
where $G_{\mu\nu}$ is an hermitian matrix defined by
\beq
G_{\mu\nu} = \{ A_\mu , A_\nu  \}.
\eeq
Integrating out the auxiliary field $Q_{\mu\nu}$,
one reproduces the original action
(\ref{bosonicaction2}).

Note that the simultaneous update of each of $(Q_{\mu\nu})_{ij}$ can
be done easily by generating Gaussian variables.  As for the update of
$A_{\mu}$, note that only the components which include at least one
common index are coupled due to the second term of
(\ref{eq:vectorize}).  The $N$ diagonal components $(A_\mu)_{ii}$
($i=1,\cdots,N$) for each $\mu$ can be updated simultaneously.  The
$\frac{N}{2}$ off-diagonal components $(A_{\mu})_{n_1 n_2}$,
$(A_{\mu})_{n_3 n_4}$, ..., $(A_{\mu})_{n_{N-1} n_N}$, where
$n_1,n_2,\cdots,n_N$ are different indices, can be updated
simultaneously for each $\mu$.

\end{document}